\documentclass[12pt]{article}
\usepackage{amssymb}

\makeatletter
\@addtoreset{equation}{section}
\makeatother

\begin{document}
\rightline{RU03--B }
\vskip 2 truecm
\centerline {\bf UNIVERSALITY OF LOW-ENERGY SCATTERING IN $2+1$ DIMENSIONS:}
\centerline {\bf THE NON SYMMETRIC CASE}
\vspace{24pt}
\baselineskip 12pt
\centerline{N.N. Khuri${}^{(a)}$, Andre Martin${}^{(a)(b)}$, Pierre C. Sabatier${}^{(c)}$ and Tai Tsun Wu${}^{(d)}$}
\vspace{0.25in}
\centerline{$^{(a)}${Department of Physics, The Rockefeller University, New
York, New York 10021-6399}}
\vspace{0.1in}
\centerline{$^{(b)}${Department of Physics, CERN, Theory Division, Geneve 23, Switzerland}}
\vspace{0.1in}
$^{(c)}$ Laboratoire de Physique Mathematique, Universite de Montpellier II\\
Sciences et Techniques du Languedoc, F-34095, Montpellier, France\\

\vspace{0.1in}
\centerline{$^{(d)}${Gordon McKay Laboratory, Harvard University, Cambridge, Massachusetts 02138-2901}}
\vspace{1in}

\begin{abstract}

For a very large class of potentials, $V(\vec{x})$, $\vec{x}\in
R^2$, we prove the universality of the low energy scattering
amplitude, $f(\vec{k}', \vec{k})$.  The result is
$f=\sqrt{\frac{\pi}{2}}(1/log~ k)+o(1)/(log\frac{1}{k})$. The only
exceptions occur if $V$ happens to have a zero energy bound state.
Our new result includes as a special subclass the case of
rotationally symmetric potentials, $V(|\vec{x}|)$.
\end{abstract}

\newpage
\hsize=6in
\hoffset=-.5in
\baselineskip=2\baselineskip
\section{Introduction}
\vspace{.25in}
\hspace{.25in}In a recent paper we proved an interesting universality property for the low-energy
scattering limit in two space dimensions.$^{1}$  This was done both for massive quantum field theory
in 2+1 dimensions, and for non-relativistic quantum mechanics in two space dimensions for a centrally
symmetric force.

The result briefly stated is that the $S$-wave phase shift, $\delta_0(k)$, $k$ being the c.m. momentum,
vanishes as $\delta_0\rightarrow c/log(k/m)$ as $k\rightarrow 0$, or in exceptional cases,
$\delta_0= O(k^2)$.  The constant, $c$, is universal: $c=\pi/2$ independent of the dynamics.  For
potential scattering this kind of universality was first noted in ref. 2, albeit with an incomplete proof which
missed among other things the exceptional class of potentials.  For the field theoretic case the result can be
found in an earlier paper.$^{3}$  But it is buried in a much more general context and its physical significance
was not discussed.

Physics in two space dimensions was initially mostly of theoretical and mathematical interest.  However,
especially in the five years since ref. 1 appeared, we have had several physical experimentally accessible systems
which have two space dimensions.  These systems appear in condensed matter physics, and a recent review is given
in ref. 4.  We also note more recent theoretical papers by Lieb and Yngvasson$^{5}$ and also by Ren $^{6}$.  It is
important to note that in the condensed matter systems the forces are often not rotationally symmetric and
in some cases they are also non-local.

In the present paper we return to the non-relativistic case but treat potentials which have no rotational symmetry,
$V\equiv V(\vec{x})$, $\vec{x}\varepsilon R^2$.  In this case there are no phase-shifts, but we obtain the
corresponding low energy result for the full amplitude which agrees with that obtained in the rotationally symmetric
case.  This is obtained under very general and "reasonable" conditions on $V(\vec{x})$ which have the following three
properties:  a.)  They are linear in $V$; b.)  They are invariant under a shift of origin, and c.)  They
include the previously studied case of ref.1.

Section II is devoted to preliminaries and definitions, including the Green's function in $R^2$ and the scattering
integral equation.

In Section III we study the Fredholm integral equation with the $2D$ zero energy Green's function,
$G_0=(\frac{1}{2\pi})log|\vec{x}-\vec{y}|$.  The main task in this section is to define the general class of
potentials, $V(\vec{x})$, to be considered.

In Section IV, we consider the Lippmann-Schwinger equation for $k>0$.  We prove that, given our class of potentials, then for
any real fixed $k$ with $k>0$, this equation has a unique solution, $\psi(\vec{k},\vec{x})$, with $\psi\varepsilon \it{C}$,
where $\it{C}$ is the Banach space of continuous functions on $R^2$ with a sup norm.  We also obtain a k-dependent upper bound for
$\parallel\psi\parallel$, which will prove useful in the succeeding sections.

Section V is devoted to proving that the zero energy kernel defines a compact operator on $C$.  We also show that even
in the case where there exist non-trivial solutions of the homogeneous integral equation with a zero energy kernel,
one still has solutions of the inhomogeneous equation.  However these are not unique.  This helps in solving the
"exceptional case" where $\phi_j, j=1 ..., N,$ are solutions of the homogeneous equation but with
$\int d^2x\phi_j(\vec{x})V(\vec{x})=0$.

We end up with two cases to consider.  Case A is where the solution(s) of the inhomogeneous equation,$\tilde{\phi}$, are such that
$\int d^3x\tilde\phi(\vec{x})V(\vec{x})\neq 0$.  Case B is where the solutions of the inhomogeneous equations satisfy
$\int d^3x\tilde\phi(\vec{x})V(\vec{x})= 0$.

In Section VI we prove the universality of the low energy
scattering amplitude for case A.  The result for the full
scattering amplitude $f(\vec{k'},\vec{k})$ is
$f=-\sqrt\frac{\pi}{2}(log\frac{1}{k})^{-1}+o(1)/(log k)$.  This
agrees with our result for the symmetric case given in ref. 1.

In Section VII, we treat the "exceptional case", i.e. case mainly
B, and obtain $f=o(1)/log(\frac{1}{k})$.  Finally in Section VIII
we discuss two additional exceptional cases, $A_{II}$ and $B_{I}$.

We briefly discuss the case of non-local potentials in Section IX.
This is limited to giving the class of non-local potentials
$W(\vec{x},\vec{y})$, that can be studied by our methods.

Finally, in the last section we give a series of comments and conclusions related to issues raised by this work.

\section{Non-Relativistic Scattering in 2D}
\vspace{.25in}
\hspace{.25in}The free Green's function in two dimensions is given by,
\begin{equation}
G(\vec{x},\vec{y})\equiv\frac{1}{4i}H^{(1)}_0(k\mid\vec{x}-\vec{y}\mid),
\end{equation}
where
\begin{equation}
(\nabla^2+k^2)G(\vec{x},\vec{y})=\delta^2(\vec{x}-\vec{y}),
\end{equation}
with $\vec{x},\vec{y}\varepsilon R^2$, and $H^{(1)}_0$ is the standard Hankel function.

The scattering integral equation is
\begin{equation}
\psi(\vec{k},\vec{x})=e^{i\vec{k}.\vec{x}}+\frac{1}{4i}\int d^2yH_0^{(1)}(k\mid\vec{x}-\vec{y}\mid)V(\vec{y})
\psi(\vec{k},\vec{y}).
\end{equation}
The class of non-central potentials, $V$, will be specified in the next section.

The asymptotic behavior of $\psi$ for large $\mid\vec{x}\mid$ is given by
\begin{equation}
\psi(\vec{k},\vec{x}){\overrightarrow{{_{\mid\vec{x}\mid\rightarrow\infty}}}}e^{i\vec{k}.\vec{x}}+\frac{i}{\sqrt{k}}
f(\vec{k}',\vec{k})\frac{e^{i(kr-\frac{\pi}{4})}}{\sqrt{r}}, ~r=|\vec{x}|, ~\vec{k}'=k\frac{\vec{x}}{|\vec{x}|}.
\end{equation}
Here we have used the large $\mid z\mid$ behavior of $H^{(1)}_o(z)$.  This leads to
\begin{equation}
G~{\overrightarrow{{_{\mid\vec{x}-\vec{y}\mid\rightarrow\infty}}}}\frac{1}{i}\sqrt\frac{1}
 {8\pi k|\vec{x}-\vec{y}|}~e^{i(k\mid\vec{x}-\vec{y}\mid-\frac{\pi}{4})}.
\end{equation}

Equation (2.4) defines the scattering amplitude, $f(\vec{k'},\vec{k})$.  One should note that in Eq. (2.4) we have chosen a
center, i.e. the point $\vec{x}=0$.  Unlike the rotationally symmetric case, the definition of $f$ is only unique
up to a phase.  Shifting the center by $\vec{a}, \vec{x}\rightarrow\vec{x}+\vec{a}$, the new amplitude differs by a
factor $exp(i(\vec{k}'-\vec{k}).\vec{a})$.    One should note that the forward scattering amplitude is invariant under
this shift as one would expect from the optical theorem.  In this paper, we are only interested in the $k\rightarrow 0$
limit, which is clearly independent of the choice of a center.

From Eqs. (2.2) and (2.4) we have the standard expression for $f$,
\begin{equation}
f(\vec{k}',\vec{k})=\frac{-1}{\sqrt{8\pi}}\int d^2x e^{-i\vec{k}'.\vec{x}}V(\vec{x})\psi(\vec{k},\vec{x}),
\end{equation}
where $\mid\vec{k'}\mid=\mid\vec{k}\mid$.

The problem we face in this paper originates from the logarithmic singularity of $H_o^{(1)}(z)$ at $z=0$.

We define $R(z)$ by the following
\begin{equation}
H^{(1)}_0(z)\equiv C_0+\frac{2i}{\pi}log~z+R(z),
\end{equation}
where
\begin{equation}
C_0 \equiv 1+\frac{2i}{\pi}[\gamma-log 2],
\end{equation}
and $\gamma$ is Euler's constant.  For small $\mid z\mid$  we have
\begin{equation}
R(z)=O(\mid z\mid^2\mid log|z||).
\end{equation}

Substituting Eq.(2.7) in (2.3) we obtain
\begin{equation}
\psi(\vec{k},\vec{x})= e^{i\vec{k}\cdot\vec{x}}+\int d^2y[\frac{log~k\mid\vec{x}-\vec{y}\mid}{2\pi}+\frac{C_0}{4i}
+\frac{R(k\mid\vec{x}-\vec{y}\mid)}{4i}] V(\vec{y})\psi(\vec{k},\vec{y}).
\end{equation}

From Eq. (2.7) we also have
\begin{equation}
R(k\mid\vec{x}\mid)=H^{(1)}_0(k\mid\vec{x}\mid)-\frac{log~k\mid\vec{x}\mid}{2\pi}-\frac{C_0}{4i}.
\end{equation}

Using this result we obtain
\begin{eqnarray}
\psi(\vec{k},\vec{x})&=&e^{i\vec{k}\cdot\vec{x}}+F(\vec{k})[\frac{H^{(1)}_0(k\mid\vec{x}\mid)}{4i}-\frac{log~|\vec{x}|}{2\pi}]\\ \nonumber
&+&\frac{1}{2\pi}\int d^2y~(log~\mid\vec{x}-\vec{y}\mid) V(\vec{y})\psi(\vec{k},\vec{y})\\ \nonumber
&-&\frac{1}{4i}\int[R(k\mid\vec{x}\mid)-R(k\mid\vec{x}-\vec{y}\mid)]V\vec{(y)}\psi(\vec{k},\vec{y}).
\end{eqnarray}
where $F$ is defined as
\begin{equation}
F(\vec{k})\equiv\int d^2yV(\vec{y})~\psi(\vec{k},\vec{y}).
\end{equation}

We stress that in going from Eq.(2.3) to (2.12) we have made no approximations.  We will show in Appendix B that the last
term in (2.12),is for small $k$, proportional to $[o(1)/|\ln k|](sup_{\vec{y}}|\psi(\vec{k},\vec{y})|)$, and is thus small
compared to the term preceding.

It is clear then from Eq. (2.12) that our first task is to study the zero energy kernel,
$K(\vec{x},\vec{y})=(\frac{1}{2\pi}log|\vec{x}-\vec{y}|)~V(\vec{y})$, which is the main term in (2.12).  This analysis
will also give us the definition of the broadest class of $V's$ that we will investigate.

\section{The Zero Energy Kernel and the Class of Potentials}

\vspace{.25in}
\hspace{.25in}In this section we consider the integral equation with the zero energy Green's function,
$G_0=\frac{1}{2\pi}log\mid\vec{x}-\vec{y}\mid$.  We use it to define our class of potentials $V(\vec{x})$,
essentially as those that lead to a kernel, $K$ which is bounded on a Banach space of continuous functions
on $R^2$.

We start with
\begin{equation}
\phi(\vec{x})= 1+\frac{1}{2\pi}\int d^2y[log|\vec{x}-\vec{y}|]V\vec{(y)}\phi(\vec{y}).
\end{equation}
It is more convenient to deal with functions that have a finite sup norm.
We define
\begin{equation}
u(\vec{x})=\frac{\phi(\vec{x})}{log(2+|\vec{x}|)},
\end{equation}
\begin{equation}
u_0(\vec{x})= \frac{1}{log(2+|\vec{x}|)}.
\end{equation}
The resulting integral equation is
\begin{equation}
u(\vec{x})=u_0(\vec{x})+\int d^2yK(\vec{x},\vec{y})u(\vec{y}),
\end{equation}
with
\begin{equation}
K(\vec{x},\vec{y})=\frac{1}{2\pi}(log|\vec{x}-\vec{y}|)V(\vec{y})\frac{log(2+|\vec{y}|)}{log(2+|\vec{x}|)}.
\end{equation}

The objective is to study Eq. (3.4) for $u\varepsilon\it{C}$, the Banach space of all bounded continuous functions on $R^2$.
The norm on $\it{C}$ is
\begin{equation}
\|u\|=\sup\limits_{\vec{x}\epsilon R^2}|u(\vec{x})|.
\end{equation}
$\it{C}$ is complete and convergence in the norm is uniform convergence.

Our task in this section is to define suitable conditions on $V(\vec{x})$ that are needed to guarantee that
$K(\vec{x},\vec{y})$ is a bounded operator on $\it{C}$, i.e. we seek some sufficient conditions on $V(\vec{x})$
such that
\begin{equation}
\int d^2y|log|\vec{x}-\vec{y}|V(\vec{y})|\frac{log(2+|\vec{y}|)}{log(2+|\vec{x}|)}< M,
\end{equation}
for all $\vec{x}\epsilon R^2$.

By suitable conditions we mean:\\
i)   The conditions are linear in $|V(\vec{x})|$;\\
ii)  The conditions are invariant under a shift of origin; \\
iii) The conditions are invariant under a scale change; and\\
iv)  The previous symmetrical case of ref. 1 is included. \\

Here iii) implies that $k$ can be replaced by any positive
multipole of $k$.  In particular, the $k$ of Sec. 2 can be
replaced by any positive number, an especially convenient choice
being 1.  To show this explicitly, note that, for $k>0$; there is
the following inequality:
\begin{equation}
\frac{2+k|\vec{x}|}{2+\vec|{x}|}\leq max(1,k).
\end{equation}
This inequality implies immediately
$$log(2+k|\vec{x}|)-log(2+|\vec{x}|)\leq max(0,log k)$$
and
\begin {equation}
log(2+k|\vec{x}|)-log(2+|\vec{x}|)\leq max(log k,log \frac{1}{k}).
\end{equation})
We prove that the following two conditions are sufficient to
guarantee the validity of Eq. (3.7) and hence the boundedness of
$K$,
\begin{equation}
~~~~~~~(A.)~~~~~~~~~~~~~~\int d^2y|V(\vec{y})|(log(2+|\vec{y}|))^2<M.
\end{equation}
and
\begin{equation}
(B.)~~~~~~~~~~~~~\int^1_0 y dy|log y|~||V(\vec{y})|_R<M,
\end{equation}
where $|V(\vec{y})|_R$ is the rearrangement of $|V(\vec{y})|$.

We remind the reader that the circular decreasing rearrangement of
a non-negative function, $f(\vec{x})$, is a decreasing function,
$f_R(|\vec{x}|)$, such that
\begin{equation}
\mu[f_R(|\vec{x}|)\geq A]=\mu[f(\vec{x})\geq A],~ \forall A;
\end{equation}
where $\mu$ is the Lebesgue measure.

For the proof, we introduce for positive $r$,
\begin{eqnarray}
log^+ r&=& max(log r,0),\\ \nonumber
log^- r&=& max(-logr,0).
\end{eqnarray}
Hence we have
\begin{eqnarray}
log r &=& log^+r - log^- r,\\ \nonumber
|log r|&=& log^+r+log^- r.
\end{eqnarray}

This splits Eq. (3.7) into two inequalities,
\begin{equation}
\int d^2y log^+|\vec{x}-\vec{y}|~|V(\vec{y})|\frac{log(2+|\vec{y}|)}{log(2+|\vec{x}|)} < M,
\end{equation}
and
\begin{equation}
\int d^2y log^-|\vec{x}-\vec{y}|~|V(\vec{y})|\frac{log(2+|\vec{y}|)}{log(2+|\vec{x}|)} < M.
\end{equation}

We consider Eq. (3.13) first.  Fixing $\vec{y}$, we have
\begin{equation}
\max_{\vec{x}}\frac{log^+|\vec{x}-\vec{y}|}{log(2+|\vec{x}|)}= \max_{\vec{x}}\frac{log |\vec{x}-\vec{y}|}{log(2+|\vec{x}|)}
= \max_{\vec{x}}\frac{log(x+y)}{log(2+x)}.
\end{equation}
But, for fixed $y$, the last expression above is monotonic in $x$.  Therefore we obtain
\begin{equation}
\max_{\vec{x}}\frac{log^+|\vec{x}-\vec{y}|}{log(2+|\vec{x}|)}\leq\frac{log(2+y)}{log 2}.
\end{equation}
Hence the inequality (3.13) is satisfied provided that the condition $(A)$, given in Eq. (3.8) is true.  One should
note that $(A)$ is invariant under a change of origin.

Next we go to the inequality (3.14).  Because of the factor $log^-|\vec{x}-\vec{y}|$, we know that the integrand in (3.13) is zero if $|\vec{x}-\vec{y}|>1$.  Hence our domain of integration is such that
\begin{equation}
x-1< y <x+1.
\end{equation}
This gives
\begin{equation}
log(2+y)<log(3+x).
\end{equation}
But since $[log(3+x)/log(2+x)]\leq log3/log2$, the validity of Eq. (3.14) reduces to
\begin{equation}
\int d^2y log^-|\vec{x}-\vec{y}|\cdot| V(\vec{y})|<\frac{log 3}{log 2}~M.
\end{equation}

For any $f(\vec{r})$ we denote by $f_R(r)$ the rearrangement of $|f(\vec{r})|$.  One has the inequality
\begin{equation}
\int d^2 y|f(\vec{y})|\cdot|g(\vec{y})|\leq\int d^2 yf_R(y)g_R(y),
\end{equation}
and using this result
\begin{eqnarray}
\int d^2y log^-|\vec{x}-\vec{y}|\cdot |V(\vec{y})|&\leq&\int d^2y (log^-|\vec{x}-\vec{y}|)_R\cdot |V(\vec{y})|_R\\ \nonumber
&\leq&\int d^2y (log^- y)_R\cdot |V(\vec{y}|)|_R\\ \nonumber
&\leq&\int d^2y (log^- y)|V(\vec{y})|_R.
\end{eqnarray}
But $log^- y=0$ for $y>1$, and thus
\begin{equation}
\int^1_0 ydy|log y||V(\vec{y})|_R\leq\frac{log 3}{log 2} M.
\end{equation}
This establishes Eq. (3.14), and completes our proof that for potentials, $V(\vec{x})$, satisfying $(A)$ and $(B)$, $K$ is a bounded operator on $C$.

In section $V$ we prove that $K$ is compact on $\it{C}$, and also prove other important properties of the zero energy integral equation.

\section{Solutions of the Lippmann-Schwinger Equation for $k>0$}
\vspace{.25in}

\hspace{.25in}We shall now proceed to prove that for our class of potentials the integral equation (2.3), for any fixed real
$k$ with $k>0$, has a unique solution, $\psi(\vec{k},\vec{x})$, with $\psi\varepsilon {\it{C}}$.
The norm of $\psi$, $\|\psi\|$, will depend on $k$, but it is bounded for any fixed $k>0$.

The integral equation (2.3) can be written as
\begin{equation}
\psi(\vec{k},\vec{x})=e^{i\vec{k}.\vec{x}}+\int d^2y\tilde{K}(k;\vec{x},\vec{y})\psi(\vec{k},\vec{y}),
\end{equation}
where
\begin{equation}
\tilde{K}(k;\vec{x},\vec{y})\equiv\frac{1}{4i}H^{(1)}_0(k|\vec{x}-\vec{y}|)V(\vec{y}).
\end{equation}

We first prove the boundedness of $\tilde{K}$.\\
\underline{Lemma 4.1}:

For any real $k, k>0$, $\tilde{K}$ is a bounded operator on $\it{C}$, with
\begin{equation}
\parallel\tilde{K}(k)\parallel<M(k)<\infty.
\end{equation}
where $M(k)=M_1[1+log\frac{1}{k}]$.\\
\underline{Proof}:

We use the bound on $H^{(1)}_0(\lambda)$,
\begin{equation}
|H^{(1)}_o(\lambda)|\leq |H^{(1)}_0(\lambda_0)|+log^+(\frac{\lambda_0}{\lambda}).
\end{equation}
This is proved in Appendix A.  Hence we have
\begin{equation}
|H^{(1)}_0(k|\vec{x}-\vec{y}|)|<|H^{(1)}_0(k_0|\vec{x}-\vec{y}|)|+log\frac{k_0}{k};
\end{equation}
For our case we can set $k_0=1$.

Next we need the following bounds on $H^{(1)}_0(|\vec{x}-\vec{y}|)$,
\begin{eqnarray}
|H^{(1)}_0(|\vec{x}-\vec{y}|)|&<&C_1+C_2|log|\vec{x}-\vec{y}||~;~|\vec{x}-\vec{y}|<1 \\ \nonumber
|H^{(1)}_0(|\vec{x}-\vec{y}|)|&<&C_3~~~;~~~|\vec{x}-\vec{y}|\geq 1.
\end{eqnarray}
Hence, for any $\chi\varepsilon \it{C}$, we get
\begin{eqnarray}
\frac{1}{4}\int|H^{(1)}_0(|\vec{x}-\vec{y}|)|~|V((\vec{y})\chi(\vec{y})|d^2y&\leq& \parallel\chi\parallel[C_1\int\limits_{|\vec{x}-\vec{y}|<1}d^2y|V(\vec{y})|\\ \nonumber
&+&C_2\int\limits_{|\vec{x}-\vec{y}|<1} d^2y|log|\vec{x}-\vec{y}|||V(\vec{y})|\\ \nonumber
&+&C_3\int\limits_{|\vec{x}-\vec{y}|>1}d^2y|V(\vec{y}),|.
\end{eqnarray}
But in the preceding section we proved that the middle integral above is bounded.

Hence we have
\begin{equation}
\parallel \tilde{K}\parallel= \sup_{\chi\epsilon C}\frac{\parallel \tilde{K}\chi\parallel}{\parallel\chi\parallel}
\leq M_1(1+log\frac{1}{k}).
\end{equation}

The theorems in textbooks for compact bounded operators on $C$ are usually given for finite domains $|\vec{x}|<\infty$.  We use
the following lemma which is a generalization of the ones in the textbooks.

\underline{Lemma 4.2}

Let $B$ be the Banach space of bounded continuous functions on $R^m(m\geq 1)$ and $B_o$ the subset of $B$ formed by functions
which tend to 0 at infinity.  $B_o$ is a closed subspace of $B$.\\

Let $Q$ denote an operator on $B$ satisfying the following three conditions:\\
(1) For any $g\varepsilon B$,
\begin{equation}
(Qg)(x)=\int q(x,y)g(y)d^my
\end{equation}
where $q$ is an $L^1$ function on $R^m \times R^m$, and\\
(2) $\int|q(x,y)|d^m y$ exists and is bounded from above by $h(x)$ where $h$ is a continuous positive
function which tends to 0 at infinity,\\
(3)  There exists a function $\eta$ on $[0,a]$ (for some $a>0)$ such that $\eta(r)\rightarrow 0$ when
$r\rightarrow 0$, and such that, for every $g\varepsilon B$ and every $(x,x')\varepsilon R^m \times R^m$,
satisfying $|x-x'|\leq a$,
\begin{equation}
|Qg(x)-Qg(x')|<||g||\eta(|x-x'|).
\end{equation}
Then $Q$ is compact from $B$ to $B$, and in fact from $B$ to $B_o$.

A proof of this lemma will be given in Appendix B.$^7$

Next we will use lemma 4.2 to prove the compactness of the operator $\tilde{K}(k)$.

\underline{Lemma 4.3}

For any fixed real $k$, $k>0$, the operator $\tilde{K}(k)$ defined by the kernel in Eq.(4.2) is compact on $\it{C}$.\\
\underline{Proof:}

From lemma (4.1) the operators, $\tilde{K}(k)$, defined by the kernel,
\begin{equation}
\tilde{K}(k;\vec{x},\vec{y})\equiv \frac{1}{4i}H^{(1)}_0(k|\vec{x}-\vec{y}|)V(\vec{y}),
\end{equation}
are, for fixed non-zero $k$, also bounded operators on $C$.  In addition for any $\vec{x}\epsilon R^2$ we have
\begin{equation}
\int d^2y|\tilde{K}(k;\vec{x},\vec{y})|\leq M<\infty,
\end{equation}
which follows from the boundedness on $C$.  But for large $|\vec{x}|$, $|\vec{x}|>>\frac{1}{k}$, we have
\begin{equation}
\int d^2y|\tilde{K}(k;\vec{x},\vec{y})|=O(\frac{1}{\sqrt{|\vec{x}|}}).
\end{equation}
This is due to the asymptotic behavior of $H^{(1)}_0(z)$.  Hence, we can always find a constant, $M(k)$, such that
\begin{equation}
\int d^2y|\tilde{K}(k;\vec{x},\vec{y})|\leq \frac{M(k)}{\sqrt{1+|\vec{x}|}},
\end{equation}
for all $\vec{x}\epsilon R^2$.

Thus $\tilde{K}_(k)$ satisfies the first condition, i.e.(4.10), of lemma 4.2 with
\begin{equation}
\tilde{h}(|\vec{x}|)=\frac{M(k)}{\sqrt{1+|\vec{x}|}}.
\end{equation}
To establish the uniform continuity of $\textit{f}(\vec{x})$, with
$f\equiv (\tilde{K}g)(\vec{x})$, we note that, as given in Eq. (2.7),
\begin{equation}
H_0^{(1)}(k|\vec{x}-\vec{y}|)=C_0+\frac{2i}{\pi}logk|\vec{x}-\vec{y}|+\textit{R}(k|\vec{x}-\vec{y}|),
\end{equation}
and uniform continuity for the kernel $log|\vec{x}-\vec{y}|V(\vec{y})$ will be established in the next section.
The same result for the operator $(RV)$ will be given in Appendix C.  Hence, the conditions of lemma 4.2 are satisfied and
therefore $\tilde{K}(k)$ is compact on $\it{C}$ for any fixed non-zero $k$.

This completes the proof of lemma 4.3.

Using the Fredholm alternative we can now assert that a unique solution, $\psi(\vec{k},\vec{x})$, of Eq. (4.1) exists for any fixed $k>0$, unless the homogeneous equation,
\begin{equation}
\psi_0(\vec{k},\vec{x})=(\tilde{K}(k)\psi_0(\vec{k}))(\vec{x}),
\end{equation}
has a nontrivial solution, $\psi_0$.  But one can easily prove that this leads to a contradiction.\\
\underline{Lemma 4.4}

For any real fixed $k$, $k>0$, there are no nontrivial solution, $\psi_0(\vec{k},\vec{x})$, of the equation
\begin{equation}
\psi_0(\vec{k},\vec{x})=\int d^2y\tilde{K}(k;\vec{x},\vec{y})\psi_0(\vec{k},\vec{y}).
\end{equation}
\underline{Proof:}

Since $\tilde{K}=( 1/4i)H^{(1)}_0(k|\vec{x}-\vec{y}|)V(\vec{y})$ we get for large $k|\vec{x}|$,
$|\vec{x}|>>\frac{1}{k}$,
\begin{equation}
\psi_0(\vec{k},\vec{x})=i e^{i(kx-\frac{\pi}{4})}[\frac{\tilde{f}_0}{\sqrt{kx}}+O(\frac{1}{(kx)^{3/2}})].
\end{equation}
where
\begin{equation}
\tilde{f}_0=\frac{-1}{\sqrt{8\pi}}\int d^2y e^{-i\vec{k}.\vec{y}}V(\vec{y})\psi_0(\vec{k},\vec{y}),
\end{equation}
with $\frac{\vec{k'}}{|\vec{k}|}=\frac{\vec{x}}{|\vec{x}|}$, and $|\vec{k'}|=|\vec{k}|=k$.

Next $\psi_o(\vec{k},\vec{x})$ satisfies the Schrodinger equation
\begin{equation}
-\nabla^2\psi_0+V\psi_0=k^2\psi_0.
\end{equation}
This leads to
\begin{equation}
\psi^*_0\nabla^2\psi_0-\psi_0\nabla^2\psi^*_0=0,
\end{equation}
since $V$ is real.  Integrating Eq. (4.22) over a large disc $A$
\begin{equation}
\int\limits_{\partial A}(\psi^*_0\vec{\nabla}\psi_0-\psi_0\vec{\nabla}\psi^*_0)\cdot d\vec{e}=0,
\end{equation}
$\vec{e}=\vec{x}/|\vec{x}|$.  We note that
\begin{equation}
\vec{\nabla}(\frac{e^{i(kx-\pi/4)}}{\sqrt{kx}})=\frac{ike^{i(kx-\frac{\pi}{4})}}{\sqrt{kx}}\cdot
\frac{\vec{x}}{|\vec{x}|}+{\frac{\vec{x}}{|\vec{x}|\sqrt{kx}}}O(\frac{1}{x}),
\end{equation}
where the $O(1/x)$ factor has no $k$ dependence.

Using the asymptotic expression (4.24) for $\psi_0$, and substituting in Eq. (4.23), we
finally obtain as $|\vec{x}|\rightarrow\infty$,
\begin{equation}
2\pi\int d\theta|\tilde{f}_0(k,\theta)|^2=0.
\end{equation}
This leads to a contradiction and completes the proof of Lemma 4.4.

In conclusion, for our class of potentials, and any $k>0$, a unique solution, $\psi(\vec{k},\vec{x})$, of Eq. (4.1) exists and is in $\it{C}$, i.e.
\begin{equation}
\parallel\psi\parallel=\sup_{\vec{x}}|\psi(\vec{k}
,\vec{x})|\leq M_1(k)<\infty.
\end{equation}
The norm $\parallel\psi\parallel$ of course depends on $k$, and in principle could grow as $k\rightarrow 0$.

\section{Compactness of the Zero Energy Operator}
\vspace{.25in}

\hspace{.25in}We consider the integral equation (3.4)
\begin{equation}
u = u_0+K u,
\end{equation}
with $u\epsilon {\it{C}}$, $u_0(x)= 1/log(2+|\vec{x}|)\}$,
and
\begin{equation}
K(\vec{x},\vec{y})=\frac{1}{2\pi}(log|\vec{x}-\vec{y}|)\frac{log(2+|\vec{y|})}{log(2+|\vec{x}|)}V(\vec{y}).
\end{equation}

It is easy to check that Lemma 4.2 does not apply to $K(\vec{x},\vec{y})$ because of the large $|\vec{x}|$ behavior of $K(\vec{x},\vec{y})$.  This difficulty can be bypassed by writing
\begin{equation}
K \equiv K^{(1)} + B,
\end{equation}
where
\begin{equation}
K^{(1)}(\vec{x},\vec{y})=\frac{1}{2\pi}\{log|\vec{x}-\vec{y}|-log(2+|\vec{x}|)\}\frac{log(2+|\vec{y}|)}
{log(2+|\vec{x}|)}V(\vec{y});
\end{equation}
and $B$ is a separable kernel
\begin{equation}
B(\vec{x},\vec{y})=\frac{1}{2\pi}(log(2+|\vec{y}|))V(\vec{y}).
\end{equation}
\underline{Lemma 5.1}

$K^{(1)}$ defines a compact operator on $\it{C}$.\\
\underline{Proof:}

The first condition of lemma 4.2 applies to $K^{(1)}$.  Indeed we have
\begin{equation}
\int|K^{(1)}(\vec{x},\vec{y})|d^2y\leq \tilde{h}(\vec{x}).
\end{equation}
Here $\tilde{h}(|\vec{x}|)=o(1)$ for large $x$, and $\tilde{h}\rightarrow 0$ as
 $|\vec{x}|\rightarrow\infty$.  This can be easily shown using the methods of Section III.

Next we have to establish uniform continuity as given in inequality (4.10) in lemma 4.2.

We have, for any $u(\vec{x})\epsilon C$, the image $w(\vec{x})$ given by
\begin{equation}
w(\vec{x})=\frac{1}{log(2+|\vec{x}|)}\int d^2y[log|\vec{x}-\vec{y}|-log(2+|\vec{x}|)]log(2+|\vec{y}|)V(\vec{y})u(\vec{y}).
\end{equation}

For any $\vec{x}_0$, $\delta>0$, we take the discs,
\begin{eqnarray}
|\vec{x}-\vec{x}_0|\leq 2\delta, \\ \nonumber
|\vec{x}'-\vec{x}_0|\geq 2\delta.
\end{eqnarray}
We want to find a uniform bound on $|w(\vec{x})-w(\vec{x}')|$ which depends only on $\delta$ and not $\vec{x'}$ or $\vec{x}$.  Clearly,
the second term in the bracket in (5.7) presents no difficulty and we only need to bound
$|\tilde{w}(\vec{x})-\tilde{w}(\vec{x}')|$ where
\begin{equation}
\tilde{w}(\vec{x})=\frac{1}{2\pi}\frac{1}{log(2+|\vec{x}|)}\int d^2y(log|\vec{x}-\vec{y}|)V(\vec{y})log(2+|\vec{y}|)u(\vec{y}).
\end{equation}

We can now write
\begin{eqnarray}
\tilde{w}(\vec{x})-\tilde{w}(\vec{x}')&=&\frac{1}{2\pi}[\frac{1}{log(2+|\vec{x}|)}-\frac{1}{log(2+|\vec{x}'|)}]\\ \nonumber
&\int& d^2y(log|\vec{x}-\vec{y}|)V(\vec{y})log(2+|\vec{y}|)u(\vec{y})\\ \nonumber
&+&\frac{1}{2\pi}\frac{1}{log2+|\vec{x}'|}\int d^2y(log|\frac{\vec{x}-\vec{y}}{\vec{x}'-\vec{y}}|)V(\vec{y})log(2+|\vec{y}|)u(\vec{y}).
\end{eqnarray}

We treat the two terms in (5.10) separately
\begin{equation}
\tilde{w}(\vec{x})-\tilde{w}(\vec{x}')\equiv I_1+I_2.
\end{equation}
It follows immediately that
\begin{equation}
|I_1|\leq \frac{|\vec{x'}|-|\vec{x}|}{(2+|\vec{z}|)[log(2+|\vec{x}|)]^2}~|\int d^2ylog|\vec{x}-\vec{y}|V(\vec{y})log(2+|\vec{y}|)u(\vec{y})|;
\end{equation}+
where we have assumed $|\vec{x}'|>|\vec{x}|$, and
\begin{equation}
|\vec{x}|<|\vec{z}|<|\vec{x}'|
\end{equation}
The integral in (5.12) is bounded for $u\epsilon \it{C}$, since it is almost identical to the integrals studied in Eqs. (3.7) and (3.8).  Thus we have
\begin{equation}
|I_1|\leq \frac{|\vec{x}-\vec{x}'|}{2 log(2+|\vec{x}|)}\cdot\frac{C}{2+|\vec{x}|}
\end{equation}

For $I_2$ we introduce two potentials $V_{x_0,\delta}(\vec{x})$ and $W_{x_0,\delta}(\vec{x})$ such that
\begin{eqnarray}
V_{x_0,\delta}&=& V, ~~~~for~~~~~|\vec{x}-\vec{x}_0|\leq 2\delta,\\ \nonumber
&=& 0,~~~~~~~~~~~~~~~~|\vec{x}-\vec{x_0}|>2\delta.
\end{eqnarray}

and
\begin{eqnarray}
W_{x_0,\delta}&=&0,~~~~~~for~~~~~|\vec{x}-\vec{x_0}|<2\delta,\\ \nonumber
&=& V,~~~for~|\vec{x}-\vec{x_0}|>2\delta.
\end{eqnarray}

Next we have
\begin{eqnarray}
|I^{(1)}_2|&\leq& \frac{\parallel
u\parallel}{log(2+|\vec{x}'|)}\int
d^2y|log\frac{|\vec{x}-\vec{y}|}{|\vec{x}'-\vec{y}|}
||W_{x_0,\delta}(\vec{y})|(log(2+|\vec{y}|))\\\nonumber &\leq&
\frac{\parallel u\parallel}{log(2+|\vec{x}'|)}\int
d^2y|[log\frac{|\vec{x}'-\vec{y}|+|\vec{x}-\vec{x}'|}{|\vec{x}'-\vec{y}|}\\
\nonumber
&+&log\frac{|\vec{x}'-\vec{y}|+|\vec{x}-\vec{x}'|}{|\vec{x}-\vec{y}|}]
|W_{x_0,\delta}(\vec{y})|(log(2+|\vec{y}|))
\end{eqnarray}
But in this integral $W_{\vec{x}_o,\delta}(\vec{y})$ vanishes for $|\vec{y}-\vec{x_o}| \leq 2\delta$.  Hence it follows from
Eq. (5.9) that our region of integration over $\vec{y}$ is such that
\begin{eqnarray}
|\vec{x}'-\vec{y}|&>&2\delta-\delta>\delta,\\ \nonumber
|\vec{x}-\vec{y}|&>&2\delta>\delta.
\end{eqnarray}
Hence
\begin{eqnarray}
|I^{(1)}_2|&\leq& \frac{C_2}{log(2+|\vec{x}'|)}\cdot\frac{|\vec{x}-\vec{x}'|}{\delta}\int d^2y|V_N(\vec{y})|log(2+|\vec{y}|)\\ \nonumber
&\leq&\frac{C_2|\vec{x}-\vec{x}'|}{\delta}||u||.
\end{eqnarray}

Next we estimate $I^{(2)}_2$
\begin{equation}
|I^{(2)}_2|\leq \frac{\parallel u\parallel}{log(2+|\vec{x}'|)}\int d^2y|log\frac{|\vec{x}-\vec{y}|}{|\vec{x}'-\vec{y}|}|~|V_{x_0,\delta}(\vec{y})|log(2+|\vec{y}|)
\end{equation}
If we take $\delta<\frac{1}{3}$, $|\vec{x}-\vec{y}|$ and $|\vec{x}'-\vec{y}|$ in this integral are both less than one,
hence
\begin{equation}
|I_2^{(2})|\leq C\int d^2y[|log|\vec{x}-\vec{y}||+|log|\vec{x}'-\vec{y}||]|V_{x_0,\delta}(\vec{y})||u||.
\end{equation}
We can replace $log|\vec{x}-\vec{y}|$ and $log|\vec{x}'-\vec{y}|$ by $log^-|\vec{x}-\vec{y}|$ and
$log^-|\vec{x}'-\vec{y}|$.  Following the same arguments as in Section III, we get
\begin{equation}
|I^{(2)}_2|\leq C_3||u||\int^1_0~y~dy|log~y|V_{x_0,\delta}\vec({y})|_R
\end{equation}
Now, $V_{x_0,\delta}(\vec{y})$ vanishes outside a disc of radius $2\delta$.  Thus $V_{x_0,\delta}(\vec{y})|_R=0$ for $y>2\delta$.
Note also that $|V_{x_0,\delta}(\vec{y})|\leq |V(\vec{x})|$ which implies
\begin{equation}
0\leq |V_{x_0,\delta}(\vec{y})|_R\leq|V(\vec{y})|_R.
\end{equation}

We finally get
\begin{equation}
|I^{(2)}_2|\leq C\int^{2\delta}_0~y~dy|log~y|~|V(\vec{y})|_R ||u||.
\end{equation}
The consequence of the above integral allows us to choose $\delta$ such that for any small $\epsilon$
\begin{equation}
\int^{2\delta}_0 y~dy|log~y|~|V(\vec{y})|_R<\epsilon.
\end{equation}
Adding up $I_1$, $|I^{(1)}_2|$ and $|I^{(2)}_2|$ we obtain
\begin{eqnarray}
|\tilde{w}(\vec{x})-\tilde{w}(\vec{x}')|&\leq& \{\frac{|\vec{x}|-|\vec{x}'|}{log(2+|\vec{x}|)(2+|\vec{x}|)}C_1+
\frac{C_2|\vec{x}-\vec{x}'|}{log(2+|\vec{x}'|)\delta}\\ \nonumber
&+&\frac{C_3\epsilon}{log(2+|\vec{x}'|)}\}||u||.
\end{eqnarray}

We can choose now $|\vec{x}-\vec{x'}|<Min(\epsilon
/\delta,\delta<1)$ and hence
\begin{equation}
|\tilde{\omega}(\vec{x})-\tilde{w}(\vec{x}')|<\frac{C_4\epsilon ||u||}{log(2+|\vec{x}|)}.
\end{equation}
This proves the uniform continuity of $\tilde{w}(\vec{x})$ and hence of $w(\vec{x})$.  Thus by lemma (4.2),
$K^{(1)}$ is compact.
The separable kernel $B$ is compact for our class of $V$.  Hence, since $K=K^{(1)}+B$, the operator, $K$, is compact on $\it{C}$.

The Fredholm alternative thus holds for $K$, i.e. either a unique solution of the inhomogeneous equation, (5.1)
exists with $u\epsilon \it{C}$, and $u_0=1/log(2+|\vec{x}|)$, or there must be at least one solution of the homogeneous
equation
\begin{equation}
u_1 = K~u_1
\end{equation}
(From the compactness it follows when we have $u_j, j=1, ..., N$ satisfying Eq. (5.28), $N$ is finite.)

Returning to the notation of Eq. (3.1) with $u=\phi/\ell n(2+|\vec{x}|)$, we have two cases to consider:\\
(I)  A unique solution for Eq. (3.1) exists
\begin{equation}
\phi = 1+K_{\ell}\phi.
\end{equation}
with
\begin{equation}
K_{\ell}=\frac{1}{2\pi}(log|\vec{x}-\vec{y}|)V(\vec{y}),
\end{equation}
and
\begin{equation}
|\phi(\vec{x})
\leq C ~log|\vec{x}|,~~{\textrm{as}}~~|\vec{x}|\rightarrow\infty, C\neq 0.
\end{equation}

Otherwise we have:\\
(II)  There exist nontrivial, linearly independent, $\phi_j(\vec{x})$, $j=1,...,N,$ such that
\begin{equation}
\phi_j = K_{\ell}\phi_j.
\end{equation}

\hspace{.25in}Case II can be divided into two subcases:
\begin{equation}
\int V(\vec{x})\phi_j(\vec{x})d^2x
=V_j\neq 0.
\end{equation}

In this case it is easy to see that if we define,
\begin{equation}
\phi_a(\vec{x})\equiv-\frac{\phi_j(x)}{V_j\cdot log k_1},
\end{equation}
then
\begin{equation}
\phi_a(\vec{x})=1+\frac{1}{2\pi}\int d^2y[log k_1|\vec{x}-\vec{y}|]V(\vec{y})\phi_a(\vec{y}),
\end{equation}
which, except for the change of scale, $1\rightarrow k_1$, is essentially the same as Eq. (5.29).

For the second subcase a finite set of $\phi_j's$ exist each satisfying the homogeneous equation (5.33),
and in addition
\begin{equation}
\int d^2xV(\vec{x})\phi_j(\vec{x})=0,~~~~j=1,...,N.
\end{equation}

In Appendix D we give a proof of the following theorem\\
\underline{Theorem 5.1}

If the homogeneous equation,
\begin{equation}
\phi(x)=\frac{1}{2\pi}\int d^2y(log|\vec{x}-\vec{y}|)V(\vec{y})\phi(\vec{y}),
\end{equation}
has non-trivial solutions, $\phi_j$ which satisfy
\begin{equation}
\int V(\vec{x})\phi_j(\vec{x})d^2x=0,~~~~~j=1,...,N.
\end{equation}
then the inhomogeneous integral equation has non-unique solutions, $\phi_a$,
\begin{equation}
\phi_a(x)=1+\frac{1}{2\pi}\int d^2y(log|\vec{x}-\vec{y}|)V(\vec{y})\phi_a(\vec{y}).
\end{equation}
where Eq. (5.38) is a necessary and sufficient condition for Eq. (5.39) to hold.

This theorem does not restrict $\phi_a(\vec{x})$ to have $\int
d^2xV(\vec{x})\phi_a(\vec{x})$ to vanish.

With this last theorem it becomes clear that we have four cases to consider: $A_I, A_{II}, B_I, B_{II}$.
These are defined as follows:

\underline{$A_I$}:~~~~~~$\phi=1+K_{\ell}\phi$ has a unique solution, and $\int d^2xV(\vec{x})\phi(\vec{x})\neq 0$.

\underline{$A_{II}$}:~~~There exist $N$ linearly independent
solutions,$\phi_j(\vec{x}), j=1, ... , N,$ for the

\hspace{.25in}homogeneous equation, $\phi_j=K_{\ell}\phi_j$. But
all the $\phi_j$ satisfy $\int V(\vec{x})\phi_j(\vec{x})d^2x\neq
0$.

\underline{$B_I$}:~~~~~$\phi=1+K_{\ell}\phi$ has a unique solution but $\int d^2xV(\vec{x})\phi(\vec{x})=0$.

\underline{$B_{II}$}:~~~The homogeneous equation,
$\phi=K_{\ell}\phi$, has $N$ linearly independent solutions,

\hspace{.25in}$\phi_j, j=1, ... ,N,$ but $\int
d^2xV(\vec{x})\phi_j(\vec{x})=0$.

One should note that the trivial case $V\equiv 0$, belongs to $B_I$.  Since, $\phi =1$, is a unique
solution of $\phi=1+K_{\ell}\phi$ when $V\equiv 0$.

One should note also that in both cases $A_I$ and $A_{II}$ we have the bound
\begin{equation}
|\phi(\vec{x})|\leq Const. ~log(2+|\vec{x}|).
\end{equation}

While in cases $B_I$, and $B_{II}$ we have stronger results.   In
Appendix E we prove that for these two cases $|\phi(\vec{x})|$ is
bounded for all $\vec{x}\epsilon R_2$, and more precisely
$|\phi|\rightarrow0$ as $|\vec{x}|\rightarrow\infty$ in the case
$B_{II}$.

We have referred to these cases as "zero energy bound states".
They are the limits, for a potential $V=g\tilde{V}$, of the
trajectories of negative energy bound states $E_n(g)$, when $g$
decreases to a critical value $g_n$, where $E_n(g_n)=0$.  For
$E<0$, the wave functions decrease exponentially, and they cannot
approach a solution growing like $log|\vec{x}|$ for $g=g_n$.

For any $\kappa>0$, and $k=i\kappa$, there is a discrete set of
couplings, $g_n(\kappa)$, such that a physical bound state exists
at $E=-\kappa^2$.  Here $(1/g_n(\kappa))$ is an eigenvalue of the
homogeneous Lippmann-Schwinger equation(2.3).  The discrete nature
of $g_n(\kappa)$ follows from the compactness of the operator in
(2.3), as does the fact that at each $E$ the degeneracy is finite.
But in addition to these general properties, two of us (A.M. and
T.T.W.) have shown$^{9}$ that there exists an explicit bound on
the number of zero energy bound states.
\section{Universality for Case $A_I~$:$\int V(\vec{x})\phi(x)d^2x\neq 0$}
\vspace{.25in}
\hspace{.25in}Without introducing any approximations we can rewrite our original integral equation for $k>0$, i.e.
Eq. (2.3), in the form of Equation (2.12)
\begin{eqnarray}
\psi(\vec{k},\vec{x})&=&e^{i\vec{k}.\vec{x}}+\textit{F}(\vec{k})[\frac{H_0^{(1)}(k|\vec{x}|)}{4i}-\frac{log|\vec{x}|}{2\pi}]\\ \nonumber
&+& K_{\ell}\psi+\Delta_R\psi,
\end{eqnarray}
where $F$ is given in Eq. (2.13),
\begin{equation}
K_{\ell}(\vec{x},\vec{y})=\frac{1}{2\pi}(log|\vec{x}-\vec{y}|)V(\vec{y}),
\end{equation}
and
\begin{equation}
\Delta_R(k;\vec{x},\vec{y})=-\frac{1}{4i}\{R(k|\vec{x}|)-R(k|\vec{x}-\vec{y}|)\}V(\vec{y}).
\end{equation}

A unique solution for Eq. (6.1), $\psi$, exists and $\psi\epsilon \it{C}$, with $k>\epsilon>0$.  This was proved in section IV, for our
class of potentials.

We introduce a new Banach space, ${\bf{\cal B}}$ with a norm given
by
\begin{equation}
\|\phi\|_{\ell}\equiv\sup\limits_{\vec{x}eR^2}|\frac{\phi(\vec{x})}{log(2+x)}|,
\end{equation}
where $\phi$ is a continuous function on $R^2$.

Our first task is to estimate the norm of the operator, $\Delta_R$
for small $k$, where $\Delta_R$ is now considered as an operator
on ${\bf{\cal B}}$,

In Appendix C, the following is proved:\\
\underline{Theorem 6.1:}

As a bounded operator on the Banach space ${\bf{\cal B}}$, we have
for $\Delta_R$
\begin{equation}
\parallel\Delta_R\parallel_{\ell}=\epsilon(k),~\epsilon(k)=o(1),~~0<k<<1.
\end{equation}

One should note at this stage the cancellations that occur in Eq. (6.1) for both
$x\rightarrow\infty$ and $\vec{x}\rightarrow 0$.  First,
the $\frac{1}{2\pi}log|\vec{x}|$ term in the bracket is, for $x\rightarrow\infty$, exactly cancelled by the
contribution
of $K_{\ell}\psi$.  Second, for $x\rightarrow0$, the $log|\vec{x}|$ in the bracket is cancelled by a
$log|\vec{x}|$
coming from $H^{(1))}_0(kx)/4i$ for small $|\vec{x}|$.

Our second task is to get a bound on $\parallel\psi(\vec{k},\vec{x})\parallel_{\ell}$.\\
\underline{Lemma 6.1:}

For any $k>0$, we have the bound
\begin{equation}
\parallel\psi(\vec{k},\vec{x})\parallel_{\ell}\leq C_1+C_2|\textit{F}(\vec{k})|[log\frac{1}{k}].
\end{equation}

\underline{Proof:}

From the integral equation
\begin{equation}
\phi=1+K_{\ell}\phi
\end{equation}
we have for case $A_I$,  with
\begin{equation}
\parallel(1-K_{\ell})^{-1}\parallel_{\ell}\leq M<\infty.
\end{equation}
We set $(1-K_{\ell})^{-1}\equiv I_{\ell}$.  The full integral equation is now
\begin{equation}
(1-K_{\ell}-\Delta_R)\psi=~e^{i\vec{k}.\vec{x}}+\textit{F}(\vec{k})B(k,x).
\end{equation}
with
\begin{equation}
B\equiv \frac{H_0^{(1)}(kx)}{4i}-\frac{log~x}{2\pi}.
\end{equation}
Hence
\begin{equation}
I_{\ell}(I_{\ell}^{-1}-\Delta_R)\psi=I_{\ell}\{e^{i\vec{k}.\vec{x}}+F(\vec{k},\vec{x})\},
\end{equation}
or
\begin{equation}
(1-I_{\ell}\Delta_R)\psi=I_{\ell}\{e^{i\vec{k}.\vec{x}}+F(
\vec{k})B\}.
\end{equation}

We finally obtain
\begin{eqnarray}
\psi&=&(1-I_{\ell}\Delta_R)^{-1}I_{\ell}\{e^{i\vec{k}.\vec{x}}+F(k)B\}\\\nonumber
&=&(I_{\ell}+I_{\ell}\Delta_R
I_{\ell})\{e^{i\vec{k}.\vec{x}}+F(k)B\}+O(\|\Delta_R\|_{\ell}^2).
\end{eqnarray}
but $\parallel I_{\ell}\parallel_{\ell}\leq M$, and
\begin{equation}
\parallel B\parallel_{\ell}\leq C_2(log\frac{1}{k})+C_0,
\end{equation}
thus
\begin{equation}
\parallel\psi\parallel_{\ell}\leq C_1+C_2(log\frac{1}{k})|F(\vec{k})|.
\end{equation}
which completes the proof.

Next we obtain the universal behavior of $\textit{F}(\vec{k})$ for small $k$.  We take, $\phi(\vec{x})$, the solution of the zero energy integral equation,
\begin{equation}
\phi=1+K_{\ell}\phi,
\end{equation}
which in the case $A_I$ is such that
\begin{equation}
V_0\equiv\int d^2xV(\vec{x})\phi(\vec{x})\neq 0,
\end{equation}
and thus $\phi(x)=O(log x)$ for large $|\vec{x}|$.  We multiply both sides of Eq. (6.1) by $\phi(\vec{x})V(\vec{x})$
and integrate over $d^2x$ obtaining,
\newpage
\begin{eqnarray}
\int\phi(x)V(\vec{x})\psi(\vec{k},\vec{x})d^2x &=&V_0+\int
d^2x(e^{i\vec{k}.\vec{x}}-1)V(\vec{x})\phi(\vec{x})\\
\nonumber
&+&\textit{F}(\vec{k})\int d^2x\phi(\vec{x})V(\vec{x})B(k,x)\\
\nonumber &+&\frac{1}{2\pi}\int d^2x\int d^2y
\phi(\vec{x})V(\vec{x})log|\vec{x}-\vec{y}|V(\vec{y})\psi(\vec{k},\vec{y})\\
\nonumber &+&\int d^2x\int
d^2y\phi(\vec{x})V(\vec{x})\Delta_R(k;\vec{x},\vec{y})\psi(\vec{k},\vec{y}),
\end{eqnarray}
where $B(k,x)$ is given by Eq. (6.10).

Using the integral equation for $\phi(\vec{x})$ we obtain
\begin{equation}
\textit{F}(\vec{k})[1-X_2]=V_0+X_1+X_3,
\end{equation}
where
\begin{eqnarray}
X_1(\vec{k})&\equiv&\int d^2x(e^{i\vec{k}.\vec{x}}-1)V(\vec{x})\phi(\vec{x}), \\ \nonumber
X_2(\vec{k})&\equiv&\int d^2xB(k,x)V(\vec{x})\phi(\vec{x}), \\ \nonumber
X_3(\vec{k})&\equiv&\int d^2x\int d^2y \phi(\vec{x})V(\vec{x})\Delta_R(k,\vec{x},\vec{y})\psi(\vec{k},\vec{y}).
\end{eqnarray}
\underline{Lemma 6.2}

The following estimates hold for small $k$,
\begin{equation}
X_1(\vec{k})=o(1)/log\frac{1}{k},
\end{equation}
\begin{equation}
X_2(\vec{k})=\frac{1}{2\pi}V_0 logk+\frac{C_0}{4i}V_0+o(1),
\end{equation}
and
\begin{equation}
|X_3(\vec{k})|=o(1)[|\textit{F}(\vec{k})|.log\frac{1}{k}+C_4]
\end{equation}
\underline{Proof:}

We define $x_0(k)$ as,
\begin{equation}
x_0(k)=\frac{1}{k(log\frac{1}{k})^p} ~~;~~p>2.
\end{equation}
From Eq. (6.20) it follows that,
\begin{eqnarray}
|X_1(k)|&<&C_1\parallel\frac{\phi(x)}{log(2+x)}\parallel[\int\limits_{x<x_0(k)}d^2x|V(\vec{x})|log(2+x)
(\frac{1}{(log\frac{1}{k})^p})]\\ \nonumber
&+&2\int\limits_{x>x_0(k)}d^2x|V(\vec{x})|~|\phi(\vec{x})|).
\end{eqnarray}
Thus,
\begin{equation}
|X_1(k)|\leq O(\frac{1}{(log\frac{1}{k})^p})+2\int\limits_{1\vec{x}|>x_0(k)}d^2x|V(\vec{x})| |\phi(\vec{x})|
\end{equation}
But, $|\phi(\vec{x})|<C\cdot log(2+|\vec{x}|)$, and we get
\begin{eqnarray}
\int\limits_{x>x_0(k)} d^2x|V(\vec{x})|~|\phi(\vec{x})|&\leq& C\int\limits_{x>x_0}d^3x|V(\vec{x})|log(2+x)\\ \nonumber
&\leq&\frac{C}{log(2+x_0(k))}\int_{x>x_0}d^2x|V(\vec{x})|[log(2+x)]^2\\ \nonumber
&\leq&\frac{o(1)}{log 1/k}.
\end{eqnarray}
Hence Eq. (6.21) holds.

The estimate of $X_2$ follows again by splitting the region of integration into two.
For $|\vec{x}|<x_0(k)$,
\begin{eqnarray}
B(\vec{k},\vec{x})&=&\frac{H_0^{(1)}(kx)}{4i}-\frac{log x}{2\pi}\\ \nonumber
&=&\frac{-1}{2\pi}~log~(\frac{1}{k})+\frac{C_0}{4i}+o(1);
\end{eqnarray}
and that gives the main part of the estimate (6.22).  The integration over the domain $|\vec{x}|>x_0(k)$ is
obviously $o(1)$.

Finally, the lemma (6.1) gives us a bound $\|\psi\|_{\ell}$ given
in Eq. (6.6). This leads to
\begin{equation}
|\psi(\vec{y})|<[C_1+C_2|(F(\vec{k})|log\frac{1}{k}]log(2+y).
\end{equation}
From Appendix C we have
\begin{equation}
|\Delta_R(\vec{x},\vec{y},\vec{k})|<C~log(1+ky)|V(\vec{y})|.
\end{equation}
Substituting these last two equations in the formula (6.20) for
$X_3(k)$, we get
\begin{equation}
|X_3(k)|\leq\tilde{C}_1[|F|log\frac{1}{k}+\tilde{C}_2]\int
d^2x\int d^2y~log(2+x)|V(x)|~|V(y)|log(1+ky)log(2+y).
\end{equation}
By splitting the $y$ integration into two regions
$|\vec{y}|<x_o(k), |\vec{y}|\geq x_o(k)$, with $x_o(k)$ given in
Eq. (6.24), one can easily show that
\begin{equation}
\int d^2y|V(\vec{y})|log(1+ky).log(2+y)=o(1).
\end{equation}
The result given in Eq. (6.23) now follows immediately and lemma
(6.2) is proved.

We now insert our estimates of the $X_j,~j=1,2,3,$ in Eq. (6.17) and obtain
\begin{eqnarray}
\textit{F}(\vec{k})&[&1-\frac{1}{2\pi}V_0log
k-\frac{C_0}{4i}V_0+o(1)]\\ \nonumber
&=&V_0+\frac{o(1)}{log(\frac{1}{k})}+o(1)[|\textit{F}(\vec{k})|log\frac{1}{k}+C_4].
\end{eqnarray}
We write
\begin{equation}
|\textit{F}|=\textit{F}e^{i(arg\textit{F})}.
\end{equation}
and obtain
\begin{equation}
\textit{F}=\frac{V_0+o(1)}{[(\frac{1}{2\pi}\cdot
V_0+o(1))log(\frac{1}{k})+(\frac{-C_0V_0}{4i}+1)+o(1)]}.
\end{equation}
Hence, we finally have for small $k$,
\begin{equation}
\textit{F}(\vec{k})=\frac{-2\pi}{logk}+(\frac{o(1)}{log(\frac{1}{k})}).
\end{equation}

The first important consequence of this result on $\textit{F}$ is
to give a bound on $\parallel\psi\parallel_{\ell}$ which is
independent of $k$ and finite.  Indeed from lemma (6.1) and the
bound result (6.36), we obtain,
\begin{equation}
\parallel\psi\parallel_{\ell}\leq M\leq\infty.
\end{equation}

The definition of the scattering amplitude, $\textit{f}(\vec{k}',\vec{k})$ is given in Eq. (2.6), and
using the definition of $\textit{F}$ is in Eq. (2.12) we have
\begin{equation}
\textit{f}(\vec{k}',\vec{k})=\frac{-1}{\sqrt{8\pi}}\textit{F}(\vec{k}')-\frac{1}{\sqrt{8\pi}}\int d^3
\vec{x}(e^{i\vec{k}'.\vec{x}}-1)V(\vec{x})\psi(\vec{k},\vec{x})
\end{equation}
But
\begin{eqnarray}
|\frac{1}{\sqrt{8\pi}}&\int&
d^2x(e^{i\vec{k}'.\vec{x}}-1)V(\vec{x})\psi(\vec{k},'\vec{x})|\\
\nonumber
&\leq&C_1\parallel\psi\parallel_{\ell}\int_{|\vec{x}|\leq
x_0(k)}d^2\vec{x}|e^{i\vec{k}'.\vec{x}}-1|~|V(\vec{x})|log(2+x)\\
\nonumber &+&2M\int_{x\geq x_0(k)}d^2x|V(\vec{x})|log(2+x)\\
\nonumber &\leq&O(\frac{1}{(log
1/k)^p})+\frac{o(1)}{(log\frac{1}{k})}~~,~~p>2.
\end{eqnarray}

This leads to
\begin{equation}
\textit{f}=\frac{-1}{\sqrt{8\pi}}~\textit{F}+\frac{o(1)}{(log k)},
\end{equation}
and hence from (6.30) our final universal result,
\begin{equation}
\textit{f}=-\sqrt\frac{\pi}{2}[\frac{1}{log(\frac{1}{k})}]+\frac{o(1)}{(log
k)}
\end{equation}
which agrees with our result for the symmetric case, (see section IX).

In closing we prove that in this case
\begin{equation}
\lim_{k\rightarrow 0}\psi(\vec{k},\vec{x})\equiv 0.
\end{equation}
To prove this we return to the rewritten integral equation (6.1),
\begin{equation}
\psi(\vec{k},\vec{x})=e^{i\vec{k}\cdot\vec{x}}+\textit{F}(\vec{k})[\frac{H_0^{(1)}(kx)}{4i}-\frac{log x}
{2\pi}]+K_{\ell}\psi+\Delta_R\psi.
\end{equation}
The last term vanishes as $k\rightarrow 0$. For small $kx$
\begin{equation}
H^{(1)}_0=\frac{2i}{\pi} logkx+C_0+O(|kx|^2(log kx)).
\end{equation}
Hence
\begin{equation}
\lim_{k\rightarrow 0}\textit{F}(\vec{k})[\frac{H^{(1)}_0(kx)}{4i}-\frac{log x}{2\pi}]=\lim_{k\rightarrow 0}
\textit{F}(\vec{k})[\frac{log k}{2\pi}],
\end{equation}
and given the universality for $\textit{F}$, i.e. Eq. (6.30) we get
\begin{equation}
\lim_{k\rightarrow 0}\textit{F}(\vec{k})[\frac{H^{(1)}_0(kx)}{4i}-\frac{log x}{2\pi}]=-1.
\end{equation}
Thus in the limit we get
\begin{equation}
\psi(0,\vec{x})=\frac{1}{2\pi}\int d^2y log|\vec{x}-\vec{y}|V(\vec{y})\psi(0,\vec{y}).
\end{equation}
But we are in the case where no homogeneous solutions exist and
(6.47) leads to a contradiction unless:
\begin{equation}
\psi(0,\vec{x})~\equiv~0.
\end{equation}

\section{Universality for Case $B_{II}$:
~$\int~d^2xV(\vec{x})\phi(x)=0$.}
\vspace{.25in}

\hspace{.25in}We consider first the case $B_{II}$.

This case, $B_{II}$, is quite exceptional.  As stressed before, if
we introduce a coupling parameter replacing $V$ by $\lambda V$ we
have this case for a discrete infinite set of coupling values,
$\lambda_q, q=1,2,...$ . The multiplicity of homogeneous solutions
for each $\lambda_q$ is finite.  This follows from the compactness
of the zero energy kernel.

In case $B_{II}$ the homogeneous zero energy equation has $N$
solutions $N\geq 1$, $\phi_j$,
\begin{equation}
\phi_j(\vec{x})=\frac{1}{2\pi}\int d^2y~log|\vec{x}-\vec{y}|~V(\vec{y})~\phi_j(\vec{y}), ~j=1, ..., N;
\end{equation}
all with
\begin{equation}
\int d^2xV(\vec{x})\phi_j(\vec{x})=0.
\end{equation}

From theorem 5.1, proved in Appendix D, we know that non-unique
solutions, $\phi_a(\vec{x})$, of the inhomogeneous equation exist,
i.e.
\begin{equation}
\phi_a(\vec{x})=1+\frac{1}{2\pi}\int d^2y(log|\vec{x}-\vec{y}|)V(\vec{y})\phi_a(\vec{y}).
\end{equation}

Here there are two possibilities,
\begin{equation}
\int d^2xV(\vec{x})\phi_a = 0,
\end{equation}
or
\begin{equation}
\int d^2xV(\vec{x})\phi_a\neq 0.
\end{equation}
We consider the case (7.4) first.

For small $k>0$, $k<1$, the solutions $\psi(\vec{k},\vec{x})$
exist and we can write Eq. (6.1) formally as
\begin{equation}
\psi=[e^{i\vec{k}.\vec{x}}+F(\vec{k})B(\vec{k},\vec{x})]+[K_{\ell}+\Delta_R]\psi.
\end{equation}
Here$B$ is given in (6.28).  $K_{\ell}$ and $\Delta_R$ are
operator on the Banach space $\cal{B}$, defined in Section VI.

We now follow a procedure analogous to that used in ref. 1, and
first introduced by Pais and Wu.  The idea is to split Eq. (7.6)
into two equations with the same kernel but different
inhomogeneous terms.  We define $\psi_{\alpha}(\vec{k},\vec{x})$
and $\psi_{\beta}(\vec{k},\vec{x})$ as follows:
\begin{equation}
\psi(\vec{k},\vec{x})\equiv\psi_{\alpha}(\vec{k},\vec{x})-\frac{F(\vec{k})}{2\pi}(log\frac{1}{k})\psi_{\beta}
(\vec{k},\vec{x}),
\end{equation}
where now we have two integral equations defining $\psi_{\alpha}$
and $\psi_{\beta}$,
\begin{equation}
\psi=e^{i\vec{k}.\vec{x}}+(K_{\ell}+\Delta_R)\psi_{\alpha},
\end{equation}
and
\begin{equation}
\psi_{\beta}=\frac{-2\pi
B(\vec{k},\vec{x})}{(log\frac{1}{k})}+(K_{\ell}+\Delta_R)\psi_{\beta}.
\end{equation}
These last two equations are equivalent to (7.6).

Note first that both inhomogeneous terms are in $\cal{B}$,
\begin{equation}
\parallel~e^{i\vec{k}.\vec{x}}\parallel_{\ell}=1/log2.
\end{equation}
and from (6.28),
\begin{equation}
\parallel\frac{2\pi
B(\vec{k},\vec{x})}{log(\frac{1}{k})}\parallel_{\ell}\leq 1.
\end{equation}

The operators $K_{\ell}$ and ${\Delta_R}$ act on $\cal{B}$, and we
have shown in section VI and appendix C, that
$\parallel\Delta_R\parallel_{\ell}=\epsilon(k)=o(1)$ as
$k\rightarrow 0$.

Thus for small $k$, we have
\begin{equation}
\parallel K_{\ell}+\Delta_R\parallel_{\ell}~=~\parallel
K_{\ell}\parallel_{\ell}+o(1),
\end{equation}
and also
\begin{equation}
\lim\limits_{k\rightarrow 0}\parallel
K_{\ell}+\Delta_R\parallel_{\ell}=\parallel
K_{\ell}\parallel_{\ell}.
\end{equation}

We can now see the power and significance of the theorem proved in
appendix D.

As an operator on $\cal{B}$, $(K_{\ell}+\Delta_R)\rightarrow
K_{\ell}$ as $k\rightarrow 0$. Thus if
$\psi_{\alpha}(\vec{k},\vec{x})$ which for $k>0$, is an element of
$\cal{B}$ remains in $\cal{B}$ as $k\rightarrow 0$, we will have
\begin{equation}
\psi_{\alpha}(0,\vec{x})=1+K_{\ell}\psi_{\alpha}(0,\vec{x}).
\end{equation}
But without the result of appendix D, this will be puzzling.
However, with those results it follows that, if
$\psi_{\alpha}(0,\vec{x})$ exists,
\begin{equation}
\psi_{\alpha}(0,\vec{x})=\phi_a(\vec{x})+\Sigma^N_{j=1}.c_j\phi_j(\vec{x}),
\end{equation}
where $\phi_a=1+K_{\ell}\phi_a$, and $\phi_j=K_{\ell}\phi_j$.  At
this stage the $c_j's$ are arbitrary, but we will sketch later how
they can be fixed by a perturbative argument.

For any $k>0$, we know that $\psi(\vec{k},\vec{x})$ exists and is
bounded for all $\vec{x}$.  It is easy to show that the same holds
for both $\psi_{\alpha}(\vec{k},\vec{x})$ and
$\psi_{\beta}(\vec{k},\vec{x})$.  Next we assert that in any
interval $0<k\leq\sigma<<1$, and with a fixed $\vec{x},
\psi,\psi_{\alpha}$ and $\psi_{\beta}$ are continuous functions of
$k$.  We now assume that $\psi(0,\vec{x})$ exists. This leads to
both $\psi_{\alpha}(0,\vec{x})$ and $\psi_{\beta}(0,\vec{x})$
being finite.

With this physical assumption, we now get
\begin{equation}
|\psi_{\alpha}(\vec{k},\vec{x})|\leq
C_{\alpha}(\vec{x})=\sup\limits_{0\leq
k\leq\sigma}|\psi_{\alpha}(\vec{k},\vec{x})|,
\end{equation}
and
\begin{equation}
|\psi_{\beta}(\vec{k},\vec{x})|\leq
C_{\beta}(\vec{x})=\sup\limits_{0\leq
k\leq\sigma}|\psi_{\beta}(\vec{k},\vec{x})|,
\end{equation}
Both $C_{\alpha}$ and $C_{\beta}$ are finite for any $\vec{x}$,
since there can be no $k_0$, $0\leq k_0\leq\sigma$, such that the
sup above is infinite.  That will lead to a contradiction with the
statements of the previous paragraph, especially continuity.

It now follows that for any $\vec{x}$
\begin{eqnarray}
lim_{k\rightarrow 0}|\nabla_R\psi_{\alpha}|&=&0,\\ \nonumber
lim_{k\rightarrow 0}|\nabla_R\psi_{\beta}|&=& 0,
\end{eqnarray}
Hence we obtain,
\begin{eqnarray}
\psi_{\alpha}(0,\vec{x})&=&\tilde{\phi_{\alpha}}(\vec{x})=\phi_a+\Sigma^N_{j=1}c_j^{(\alpha)}\phi_j,\\
 \nonumber
\psi_{\beta}(0,\vec{x})&=&\tilde{\phi_{\beta}}(\vec{x})=\phi_a+\Sigma^N_{j=1}c_j^{(\beta)}\phi_j
\end{eqnarray}

From Appendix E we have the result that both
$\tilde{\phi_{\alpha}}$ and $\tilde{\phi_{\beta}}$ are bounded by
constants for all $\vec{x}$ including $\vec{x}\rightarrow\infty$.
This fact plus continuity leads to the result that both
$C_{\alpha}(\vec{x})$ and $C_{\beta}(\vec{x})$ in Eqs. (7.16) and
(7.17) are bounded for all $\vec{x}$.  Thus for a closed interval
$0\leq k\leq\sigma<<1$, we have
\begin{eqnarray}
|\psi_{\alpha}(\vec{k},\vec{x})|&\leq&\tilde{C}_{\alpha}=\sup\limits_{\vec{x}\epsilon R^2}C_{\alpha}(\vec{x}),\\
\nonumber
|\psi_{\beta}(\vec{k},\vec{x})|&\leq&\tilde{C}_{\beta}=\sup\limits_{\vec{x}\epsilon
R^2}C_{\beta}(\vec{x})
\end{eqnarray}
From Eq. (7.7) we have
\begin{equation}
F(\vec{k})=\int
d^2xV\psi_{\alpha}(\vec{k},\vec{x})~-~\frac{1}{2\pi}(log\frac{1}{k})F(\vec{k})\int
d^2xV(\vec{x})\psi_{\beta}(\vec{k},\vec{x}).
\end{equation}
Denoting the integrals above by $F_{\alpha}$ and $F_{\beta}$
respectively, we get
\begin{equation}
F(\vec{k})=\frac{F_{\alpha}(\vec{k})}{1+\frac{1}{2\pi}(log\frac{1}{k})F_{\beta}(\vec{k})}.
\end{equation}

We now multiply both sides of the Eq. (7.8), i.e. the
$\alpha$-equation, by $\tilde{\phi_{\alpha}}V$, and integrate.
After using the fact that
$\tilde{\phi_{\alpha}}=1+K_{\ell}\tilde{\phi_{\alpha}}$, and that
$\int~V(\vec{x})\tilde{\phi_{\alpha}}(\vec{x})d^2x=0$,we obtain as
before
\begin{equation}
F_{\alpha}(\vec{k})=\int(e^{i\vec{k}.\vec{x}}-1)V(\vec{x})\tilde{\phi}(\vec{x})d^2x+Y_{\alpha}(\vec{k}),
\end{equation}
with
\begin{equation}
Y_{\alpha}(\vec{k})=\int d^2x\int
d^2y\tilde{\phi}(\vec{x})\Delta_R(\vec{x},\vec{y},\vec{k})\psi_{\alpha}(\vec{k},\vec{y}).
\end{equation}
Given the fact that $|\tilde{\phi_{\alpha}}|$ is bounded (Appendix
E) we get
\begin{equation}
|\int
d^2x(e^{i\vec{k}.\vec{x}}-1)V(\vec{x})\tilde{\phi}(\vec{x})|=\frac{o(1)}{(log\frac{1}{k})^2}.
\end{equation}
Using the bound given in Appendix C,
\begin{equation}
|\Delta_R(\vec{x},\vec{y},\vec{k})|\leq C|V(\vec{y})|.log(1+ky),
\end{equation}
and the fact that $|\psi_{\alpha}(k,\vec{x})|$ is bounded for
small $k$, we get
\begin{eqnarray}
Y_{\alpha}(\vec{k})&\leq&C\int d^2x \int d^2y
|V(\vec{x})|~|V(\vec{y})|~log(1+ky)\\ \nonumber &\leq&C'\int
d^2y~|V(\vec{y})|~log(1+ky)=\frac{o(1)} {(log\frac{1}{k})}.
\end{eqnarray}
The last equality we get by subdividing the $d^2y$ integration
into $y<x_o(k)$, and $y>x_o(k)$ as done previously.  We finally
obtain as $k\rightarrow 0$
\begin{equation}
F_{\alpha}(\vec{k})=\frac{o(1)}{(log\frac{1}{k})}.
\end{equation}

Next we apply the same trick to Eq. (7.9).  We obtain
\begin{equation}
F_{\beta}=\frac{-2\pi}{(log\frac{1}{k})}\int
d^2x\tilde{\phi_{\beta}}(\vec{x})V(\vec{x})B(\vec{k},\vec{x})+Y_{\beta}(\vec{k})
\end{equation}
where $Y_{\beta}$ is given by Eq. (7.24) with
$\psi_{\alpha}\rightarrow\psi_{\beta}$.

The estimate of the first integral in Eq. (7.29) is given by
\begin{eqnarray}
\frac{2\pi}{(log\frac{1}{k})}&\int&
d^2x\tilde{\phi}_{\beta}(\vec{x})V(\vec{x})B(\vec{k},(\vec{x})\\
\nonumber &=& \frac{2\pi}{(log\frac{1}{k})}\int
d^2x\tilde{\phi}_{\beta}(\vec{x})V(\vec{x})[B(\vec{k},\vec{x})
+\frac{1}{2\pi}log\frac{1}{k}-\frac{C_0}{4i}],
\end{eqnarray}
But from Eq. (6.28) we get
\begin{equation}
[B(\vec{k},\vec{x})+\frac{1}{2\pi}(log\frac{1}{k})-\frac{C_0}{4i}]=o(1),~~|\vec{x}|<x_0(k),
\end{equation}
where $x_0(k)$ is given in Eq. (6.24).  On the other hand for
$|\vec{x}|>x_0(k)$, we have from (6.28)
\begin{equation}
|B(\vec{k},\vec{x})+\frac{1}{2\pi}log \frac{1}{k}|<~C log(2+x).
\end{equation}
Splitting the integration domain in (7.30) we
finally obtain
\begin{equation}
\frac{2\pi}{(log\frac{1}{k})}\int
d^2x\tilde{\phi}_{\beta}(\vec{x})V(\vec{x})B(\vec{k},\vec{x})=\frac{o(1)}{(log\frac{1}{k})}
\end{equation}

Hence we get
\begin{equation}
F_{\beta}=\frac{o(1)}{(log\frac{1}{k})}+Y_{\beta}(\vec{k}).
\end{equation}
But again as in Eq. (7.27) we have
\begin{equation}
Y_{\beta}=\frac{o(1)}{(log\frac{1}{k})},
\end{equation}
and hence finally
\begin{equation}
F_{\beta}=\frac{o(1)}{(log\frac{1}{k})}, ~~as~k\rightarrow 0.
\end{equation}
Substituting our results for $F_{\alpha}$ and $F_{\beta}$ in Eq.
(7.22) we get
\begin{equation} F=\frac{o(1)}{(log\frac{1}{k})},
~~as~k\rightarrow 0,
\end{equation}
and a similar result holds for $f$.

We are left with the case of Eq. (7.5), when
\begin{equation}
V_a\equiv\int d^2x\phi_a(\vec{x})V(\vec{x})\neq 0.
\end{equation}
It is easy to see that in this case we obtain the result of
section VI.

Using the same method as above, we get as $k\rightarrow 0$
\begin{eqnarray}
F_{\alpha}&=&~V_a+o(1), \\ \nonumber F_{\beta}&=&~V_a+o(1),
\end{eqnarray}
and finally
\begin{eqnarray}
F(\vec{k})&=&\frac{V_a+o(1)}{1+\frac{V_a}{2\pi}log(\frac{1}{k})}\\
\nonumber
&=&\frac{2\pi}{(log\frac{1}{k})}+\frac{o(1)}{(log\frac{1}{k})}.
\end{eqnarray}
This is the same leading term as in the standard case of section
VI.

In closing this section we sketch how the coefficients
$c_j^{(\alpha)}$ and $c^{(\beta)}_j$ can be determined.

We write Eq. (7.8) as
\begin{equation}
\psi_{\alpha}=f_{\alpha}+K_{\ell}\psi_{\alpha},
\end{equation}
where
\begin{equation}
f_{\alpha}\equiv e^{i\vec{k}.\vec{x}}+\Delta_R\psi_{\alpha}.
\end{equation}
From Appendix D we see that a necessary and sufficient condition
for (7.42) to have a solution is
\begin{equation}
\int\phi_j(\vec{x})V(\vec{x})f_{\alpha}(\vec{x})d^2x=0,~~j=1,...,N.
\end{equation}
Setting
$\psi_{\alpha}=\tilde{\phi_{\alpha}}+o(1)=\phi_a+\Sigma^N_{k=1}c^{(\alpha)}_k\phi_k+o(1)$
in Eq. (7.43), and using Eq. (7.44) we obtain
\begin{equation}
\Sigma^N_{k=1}A_{jk}c_k^{(\alpha)}=\omega_j+\int
d^2x\phi_j(\vec{x})V(\vec{x})(e^{i\vec{k}.\vec{x}}-1),
\end{equation}
with
\begin{eqnarray}
A_{jk}&\equiv&-\int\phi_j(\vec{x})V(\vec{x})[\Delta_R\phi_k](\vec{x})d^2x,\\
\nonumber \omega_j&=&
\int\phi_j(\vec{x})V(\vec{x})[\Delta_R\phi_a](\vec{x})d^2x.
\end{eqnarray}

The integral in (7.44) is $o(1)/(log\frac{1}{k})^2$, while both
$A_{jk}$ and $\omega_j$ are $O(\|\Delta_R\|_{\ell})=o(1)$.  Thus
to first order in $\|\Delta_R\|_{\ell}$, the degeneracy can be
removed if the matrix $A_{jk}$ has an inverse, and
$A^{-1}\vec{\omega}$ will then give $c^{(\alpha)}_k$ for
$k=1,...,N$.

\section{The Cases $A_{II}$ and $B_I$.}
\vspace{.25in}

We recall that under $A_{II}$, we have $N$ solutions, $\phi_j,
j=1,...,N$, of the homogeneous equation $\phi_j=K_{\ell}\phi_j$,
with $\int d^2x V(\vec{x})\phi_j(\vec{x})\neq 0$.  This case
depends critically on whether $N=1$, or $N\geq 2$.

For $N=1$, we can carry out a re scaling of $\phi_j$, and define
$\hat{\phi}_1$ as,
\begin{equation}
\hat{\phi}_1=-\frac{2\pi\phi_1}{V_1 log k_1},~~ V_1\equiv \int d^2
xV(\vec{x})\phi_1(\vec{x}).
\end{equation}

Then $\hat{\phi}_1(\vec{x})$ is a solution of the inhomogeneous
equation,
\begin{equation}
\hat{\phi}_1(\vec{x})=1+\frac{1}{2\pi}\int d^2y(log
k_1|\vec{x}-\vec{y}|)V(\vec{y})\hat{\phi}(\vec{y}).
\end{equation}
Hence, for $N=1$, $A_{II}$ will lead to the same result for
$f(\vec{k}',\vec{k})$ that was obtained in section VI for the case
$A_I$.

However, for $N\geq 2$, we can always take linear combinations of
the $\phi_j's$, such that
\begin{equation}
\int d^2x V(\vec{x})(\sum_{j=1}^N b_j\phi_j)=0.
\end{equation}
This reduces $A_{II}$ for $N\geq 2$ to the case $B_{II}$ treated
in section VII.

In the case $B_I$ one has a unique solution of the inhomogeneous
equation, $\phi=1+K_{\ell}\phi$, but with $\int
d^2V(\vec{x})\phi(\vec{x})=0$.  In Appendix E we show that in this
case $|\phi(\vec{x})|$ is bounded for all $\vec{x}$.  The result
for $F$ can now be obtained by setting $V_o=0$ in Eq. (6.35), and
noting that since $|\phi|$ is bounded
\begin{equation}
X_1(k)=\int
d^2x(e^{i\vec{k}.\vec{x}}-1)V(\vec{x}\phi(\vec{x})=\frac{o(1)}{(log\frac{1}{k})^2},
\end{equation}
with an additional power of $(log\frac{1}{k})$ than in Eq. (6.21).

In addition in this case we have
\begin{equation}
x_3(k)=o(1)/log(\frac{1}{k}).
 \end{equation}
 The final result for $F$ is then
 \begin{equation}
 F(\vec{k})=\frac{o(1)/(log\frac{1}{k})^2}{[1+o(1)]}=\frac{o(1)}{(log\frac{1}{k})^2}.
 \end{equation}

\section{Non-Local Potentials.}
\vspace{.25in}
\hspace{.25in}  This is the case where the interaction term in the Schrodinger equation is of the form:
$\int d^3yW(\vec{x},\vec{y})\psi(\vec{y})$, replacing the standard local term, $V(\vec{x})\psi(\vec{x})$.

Due to the length of this paper we shall only deal now with the definition of the class of non-local potentials,
$W(\vec{x},\vec{y})$.  More detailed results will be given elsewhere.

The zero energy integral equation in this case is
\begin{equation}
\phi(\vec{x})=1+\frac{1}{2\pi}\int d^2y\int d^2z(log|\vec{x}-\vec{y}|)W(\vec{y},\vec{z})\phi(\vec{z}),
\end{equation}
where the norm is given by
\begin{equation}
|\phi|= \sup_{\vec{x}\varepsilon R^2}\frac{|\phi(\vec{x})|}{log(2+|\vec{x}|)}.
\end{equation}
As in the local case we write
\begin{equation}
u\equiv\frac{\phi(\vec{x})|}{log(2+|\vec{x}|)},
\end{equation}
and obtain
\begin{equation}
u(\vec{x})=u_o(\vec{x})+\frac{1}{(2\pi)log(2+|\vec{x}|)}
\int d^2y\int d^2z log|\vec{x}-\vec{y}|W(\vec{y},\vec{z}) log(2+|\vec{z}|)u(\vec{z}|.
\end{equation}

We need conditions on $W$ to guarantee the boundedness of the double integral,
\begin{equation}
I=\frac{1}{log(2+|\vec{x}|)}\int d^2y\int d^2z(log|\vec{x}-\vec{y}|)W(\vec{y},\vec{z})(log(2+|\vec{z}|))u(\vec{z}).
\end{equation}

Using again $log A=log^+A -log^- A$, we write for $A=|\vec{x}-\vec{y}|$,
\begin{equation}
I = I^+ - I^-.
\end{equation}
We now have
\begin{equation}
|I^+|\leq \frac{\parallel u\parallel}{log(2+|\vec{x}|)}\int d^2y\int d^2 z log^+|\vec{x}-\vec{y}||W(\vec{y},\vec{z})
|log (2+|\vec{z}|).
\end{equation}
But
\begin{equation}
log^+|\vec{x}-\vec{y}|\leq log(2+(|\vec{x}|)+log(2+|\vec{y}|).
\end{equation}
Hence
\begin{eqnarray}
|I^+ |&\leq& \parallel u\parallel \int d^2y\int d^2z|W(\vec{y},\vec{z})|log(2+|\vec{z}|)\\ \nonumber
&+&\frac{\parallel u\parallel}{log(2+|\vec{x}|)}\int d^2 y\int d^2z~log(2+|\vec{y}|)log(2+|\vec{z}|)|W(\vec{y},\vec{z})|.
\end{eqnarray}

This leads to our first condition on $W$, namely
\begin{equation}
(A) ~~~~~~~~\int d^2y\int d^2z~log(2+|\vec{y}|).log(2+|\vec{z}|)|W(\vec{y},\vec{z})|<\infty.
\end{equation}

For $I^-$ we have
\begin{equation}
|I^-|\leq\frac{\parallel u\parallel}{log(2+|\vec{x}|)}\int d^2y\int d^2z|log^-|\vec{x}-\vec{y}|||W(\vec{y},\vec{z})|log(2+|\vec{z}|)
\end{equation}

Next one uses the inequality,
\begin{equation}
log(2+|\vec{z}|)\leq~log(2+|\vec{y}|)+log(1+|\vec{y}-\vec{z}|),
\end{equation}
and notes that when $|\vec{x}-\vec{y}|>1, ~log^-|\vec{x}-\vec{y}|=0$.  This allows us to write
$log(2+|\vec{y}|)\leq C~log(2+|\vec{x}|)$ in equation (8.12) when substituted in (8.11).  We obtain
\begin{eqnarray}
|I^-|&\leq& \frac{\parallel u\parallel}{log(2+|\vec{x}|)}\int d^2y\int d^2z|log^-|\vec{x}-\vec{y}||W(\vec{y},\vec{z})
|log[1+|\vec{y}-\vec{z}|]\\ \nonumber
&+&C\parallel u\parallel \int d^2y \int d^2z|log^-|\vec{x}-\vec{y}|| |W(\vec{y},\vec{z})|.
\end{eqnarray}

We define two decreasing rearrangements
\begin{equation}
R^{(1)}_W(y)\equiv [\int d^2z|W(\vec{y},\vec{z})|log(1+|\vec{y}-\vec{z}|)]_R,
\end{equation}
and
\begin{equation}
R^{(2)}_W(y)\equiv [\int d^2z|W(\vec{y},\vec{z})|]_R.
\end{equation}

This leads us to two condition on $R^{(1)}$ and $R^{(2)}$, namely
\begin{equation}
B^{(1)}: ~~~~~~~~~~\int^1_o y dy|log y|R^{(1)}_W(y)<\infty
\end{equation}
and
\begin{equation}
B^{(1)}: ~~~~~~~~~~\int^1_o y dy|log y|R^{(2)}_W(y)<\infty
\end{equation}
\section{Miscellaneous Remarks.}
\vspace{.25in}
\hspace{.25in}1.  In ref. 1, the next to leading term for the low
energy behavior of the phase-shift, $\delta_o(k)$, was given as
$O(k^2)$, i.e. $\delta_o=\frac{\pi}{2}(log k)^{-1}+O(k^2)$.  This
is true for massive relativistic field theories.  It is also
certainly true in non-relativistic potential scattering for
rotationally symmetric potentials that are $O(e^{-\mu r})$ for
large $r$ with some $\mu>0$.  For potentials that saturate the
condition $\int^{\infty}_1 rdr|V(r)|(log r)^2<\infty$, the
$O(k^2)$ above should be replaced by $o(1)/log\frac{1}{k}$.  This
will remove the inconsistency between ref. 1 and the present
paper.

A similar remark holds for the results in the exceptional case where $\delta_o= O(k^2)$ is only necessarily
true for massive or exponentially decreasing potentials.  Otherwise one has $\delta_o=o(1)/log k$ as in the present
case for the full $f$.

2.  This paper could be significantly shortened and simplified if we are willing to strengthen the condition(A) on
$V(\vec{x})$ given by $\int d^2x|V(\vec{x})|\{log(2+|\vec{x}|)\}^2<\infty$.  Even changing the power of the log from
2 to $2+\epsilon$ will simplify the proof somewhat.  We are however convinced that this condition is the
critical one, and in a certain sense we have the optimal result.  The difference between $V's$ for which the $(log)^2$
integral is convergent and those for which it diverges is apparent in our paper on the number of bound states.$^8$

3.  Finally, and indirectly related to the above remark, we have to answer the question why we chose to work on a Banach
space of wave functions, $\phi$, instead of working on a Hilbert space where the elements of the space are
$\sqrt{\tilde{V}\phi}$, and $\tilde{V}=V$ where $V(\vec{x})>0$, and $\tilde{V}=-V$ otherwise.  In fact it can be shown
 that the non-linear condition, introduced and studied in detail by one of us (P.C.S.)
 \begin{equation}
 \int d^2x\int d^2y|V(\vec{x})|(log|\vec{x}-\vec{y}|)^2|V(\vec{y})|<\infty,
 \end{equation}
 which gives an $L_2$ kernel for the zero energy integral equation follows from our linear conditions $(A)$ and $(B)$
 given in section III, i.e. $\int d^2x|V(\vec{x})|[log(2+|\vec{x}|)]^2<\infty$, and $\int d^2|V(\vec{x})|_R(log^-|\vec{x}|)<\infty$.
 Thus $(A)+(B)$ gives a smaller class of potentials.$^9$

 The reason we use the Banach space approach is also apparent if one reads section VII dealing with the exceptional
 case, and especially Appendix D.\\
\vspace{.25in}
\appendix
\section*{Appendix A}\def\thesection{A}\setcounter{equation}{0}

In this appendix we prove the bound,
\begin{equation}
|H_0^{(1)}(x)|<|H^{(1)}_0(x_0)|+log^+(\frac{x_0}{x}).
\end{equation}

First we prove
\begin{equation}
|H^{(1)}_0(x)|<\sqrt\frac{2}{\pi}(\frac{1}{\sqrt{x}}).
\end{equation}

From Nicholson's formula, (ref. 10),
\begin{equation}
|H^{(1)}_o(x)|^2=\frac{8}{\pi^2}\int^{\infty}_0K_0(2x~sinh~t)dt,
\end{equation}
we have
\begin{eqnarray}
|H^{(1)}_0(x)|^2&<&\frac{8}{\pi^2}\int^{\infty}_0K_0(2x~sinh~t)cosh~t~dt\\ \nonumber
&=&\frac{8}{\pi^2}(\frac{1}{2x})\int^{\infty}_0K_0(u) du.
\end{eqnarray}
But$^{11}$ $\int^{\infty}_0K_0(u)du=\pi/2$, and this proves the inequality (A.2).

Now using the notation of Abramovitz and Stegun,$^{12}$
\begin{equation}
|H^{(1)}_0(x)|\equiv M_0(x),
\end{equation}

$M_o$ satisfies a non-linear differential equation,
\begin{equation}
x^2M_0''+xM'_0+x^2M_0-\frac{4}{\pi^2M^3_0}=0.
\end{equation}

From $(A-2)$ it follows that
\begin{equation}
x^2M_0-\frac{4}{\pi^2M^3_0}<0.
\end{equation}
Hence,
\begin{equation}
x^2M''_0+xM'_0>0,
\end{equation}
or
\begin{equation}
\frac{d}{dx}(xM'_0)>0.
\end{equation}

This leads to
\begin{equation}
xM'_0(x)\geq\lim_{y\rightarrow 0}(yM'_0(y))=-1,
\end{equation}
since $M_0(x)\sim -log x$ as $x\rightarrow 0$.

For $0<x<x_0$,
\begin{eqnarray}
M_0(x_0)-M_0(x)&=&-\int^x_{x_0}M'_0(y)dy\\ \nonumber
&=&-\int^x_{x_0}(yM'_0(y))\frac{dy}{y}>-\int^x_{x_0}\frac{dy}{y}=-log\frac{x_0}{x}.
\end{eqnarray}

Thus for $x<x_o$,
\begin{equation}
|H^{(1)}_0(x)|<|H^{(1)}_0(x_0)|+log\frac{x_0}{x}.
\end{equation}
But $|H^{(1)}_0(x)|$ is decreasing.  This follows from Nicholson's formula since $K_0$ is also decreasing.
Thus we finally obtain the result (A-1).

\vspace{.25in}
\section*{Appendix B}\def\thesection{B}\setcounter{equation}{0}

A proof of lemma 4.2 follows.

We use Ascoli's Theorem:$^{7}$  Any bounded equicontinuous set of functions on a compact metric space is
relatively compact in the sup norm topology.

Since $\textbf{B}$ is a metric space, it suffices to prove that any
sequence $\{g_n\}_{n=1,2,...}$ in $\textbf{B}$ satisfying $||g_n||\leq M~ \forall n$ contains a subsequence
$\{g{_{n_{k}}}\}_{k=1,2,...}$ such that $\{Qg_{n_{k}}\}_{k=1,2, ...}$ converges in norm, i.e. such that, for
every integer $\ell\leq 1$ there exists a constant $L_{\ell}\geq 0$ such that
\begin{equation}
k,k' \geq L_{\ell} \Rightarrow ||Qg_{n_{k}}-Qg_{n_{k'}}||< 2^{\ell}.
\end{equation}
Let $\{g_n\}_{n=1,2,...}$ be a sequence in $\textbf{B}$ satisfying $||g_n|| \leq M$ for every $n$.  Then
$|Qg_n(x)|\leq M h(x)$ for all $n$ and all $x\epsilon\textbf{R}^m$.

Let $R_1 >0$ be such that $|x|\geq R_1\Rightarrow Mh(x)< 1/8$.  Since the ball
$B_1 =\{x\epsilon \textbf{R}^m:|x|\leq R_1\}$ is compact, and the sequence of the restrictions of the
$Qg_n$ to $B_1$ are equicontinuous, there exists a subsequence $\{g_n^{(1)}\}$ of $\{g_n\}$ and
a constant $L_1 > 0$ such that
\begin{equation}
n, n' \geq L_1\Rightarrow |Qg_n^{(1)}(x)-Qg_{n'}^{(1)}(x)|< 1/4~~ \forall x\epsilon B_1,
\end{equation}
and hence
\begin{equation}
n, n' \geq L_1\Rightarrow |Qg_n^{(1)}(x)-Qg_{n'}^{(1)}(x)|< 1/2~~ \forall x.
\end{equation}
Suppose we have defined, for every integer $p~\epsilon[1,\ell-1]$, a subsequence $\{g_n^{(p)}\}$ of $\{g_n\}$
and a constant $L_p > 0$ such that:\\
if $b >1$, $\{g_n^{(p)}\}$ is a subsequence of $\{g_n^{(p-1)}\}$, and for all $p$
\begin{equation}
n, n'\geq L_p \Rightarrow |Qg_n^{(p)}(x)-Qg_{n'}^{(p)}(x)|<2^{-p}~\forall x.
\end{equation}
We can then define $\{g_n^{\ell}\}$ in the same way as in the first step:  let $R_{\ell}$ be such that
$|x|\geq R_{\ell}\Rightarrow Mh(x)< 1/2^{2+\ell}$, and $B_{\ell}=\{x\epsilon \textbf{R}^m: |x|\leq R_{\ell}\}$.
Let $\{g_n^{(\ell)}\}$ be a subsequence of $\{g_n^{\ell-1}\}$ which converges uniformly on $B_{\ell}$,
and $L_{\ell}$ be a constant such that
\begin{equation}
n, n'\geq L_{\ell} \Rightarrow |Qg_n^{(\ell)}(x)-Qg_{n'}^{(\ell)}(x)|< 1/2^{2+\ell}~\forall x\epsilon B_{\ell},
\end{equation}
and hence
\begin{equation}
n, n'\geq L_{\ell} \Rightarrow |Qg_n^{(\ell)}(x)-Qg_{n'}^{(\ell)}(x)|< 1/2^{\ell}~\forall x
\end{equation}
The sequence $\{Qg_n^{(n)}\}$ is uniformly convergent on the whole space.

\vspace{.25in}
\section*{Appendix C}\def\thesection{C}\setcounter{equation}{0}

In this appendix we give a proof of Theorem 6.1 regarding the norm of the operator $\Delta_R$, where
\begin{equation}
\Delta_R(k;\vec{x},\vec{y})\equiv-\frac{1}{4i}\{R(k|\vec{x}|)-R(k|\vec{x}-\vec{y}|)\}V(\vec{y}),
\end{equation}
and $R(z)$ is defined by
\begin{equation}
R(z)=H_0^{(1)}(z)-C_0-\frac{2i}{\pi}log z,
\end{equation}
with $C_0$ given in Eq. (2.8).  We want to prove that, as an
operator on the Banach space $\it{B}, \parallel
\Delta_R\parallel_{\ell}$ has a bound for small $k$, $0<k<<1$,
\begin{equation}
\parallel\Delta_R\parallel_{\ell}=o(1).
\end{equation}

At the end of this appendix, in lemma C.1, we will prove the following inequality:
\begin{equation}
|R(k|\vec{x}|)-R(k|\vec{x}-\vec{y}|)|<C_1 log(1+k|\vec{y}|).
\end{equation}

From Eq. (C.1), we have
\begin{equation}
\parallel\Delta_R\parallel_{\ell}\leq \sup_{x}\int d^2y |R(k|\vec{x}|)-R(k(|\vec{x}-\vec{y}|)||V(\vec{y})|log(2+y).
\end{equation}
Using (C.4) we obtain
\begin{equation}
\parallel\Delta_R\parallel_{\ell}\leq C_1\int d^2y log(1+k|\vec{y}|)|V(\vec{y})|log(2+y).
\end{equation}

We set
\begin{equation}
y_0(k)=\frac{1}{k(log k)^2},
\end{equation}
and write
\begin{eqnarray}
\parallel\Delta_R\parallel_{\ell}&\leq& C_1\{ \int\limits_{|\vec{y}|<y_0}d^2y log(1+k|\vec{y}|)|V(\vec{y})|log(2+y)\\ \nonumber
&+&\int\limits_{|\vec{y}|>y_0} d^2y
log(1+k|\vec{y}|)|V(\vec{y})|log(2+y)\}.
\end{eqnarray}
This leads to
\begin{eqnarray}
\parallel\Delta_R\parallel_{\ell}&\leq& C_1\{\frac{1}{(log k)^2}\int\limits_{|\vec{y}|<y_0}|V(\vec{y})|d^2y log(2+y)\\ \nonumber
&+&C_2\int\limits_{|\vec{y}|>y_0}[log(2+|\vec{y}|)]^2|V(y)|d^2y
\end{eqnarray}
The last integral is convergent over all of $R^2$ and $y_0(k)\rightarrow\infty$ as $k\rightarrow 0$, we finally obtain
\begin{equation}
\parallel\Delta_R\parallel_{\ell}\leq~o(1),
\end{equation}
which proves Theorem (6.1).

This result is, in a certain sense, optimal since if we take the special case $\vec{x}=0$, we get for any
$\psi\varepsilon \it{C}$
\begin{equation}
(\Delta_R\psi)(0)=\frac{-1}{4i}\int d^2y R(k|\vec{y}|)V(\vec{y})\psi(\vec{y}).
\end{equation}
By taking a special $V(\vec{y})$
\begin{eqnarray}
|V(\vec{y})|&=&0, y\leq 1\\ \nonumber
|V(\vec{y})|&=&\frac{1}{y^2|log y|^{3+\varepsilon}},~~y>1.
\end{eqnarray}
We can easily show that
\begin{equation}
|(\Delta_R\psi)(0)|=O(\frac{1}{(log \frac{1}{k})^{1+\epsilon}}).\parallel \psi\parallel,
\end{equation}
for any $\epsilon> 0$.

We now prove Eq. (C.4),\\
\underline{Lemma C.1}:
\begin{equation}
|R(k|\vec{x}|)-R(k|\vec{x}-\vec{y}|)|<C log(1+k|\vec{y}|).
\end{equation}

\underline{Proof:}

The bound given in Eq. (A.2) for $|H_0^{(1)}(z)|$ and the definition of $R(z)$ given in (B.2) lead us
immediately to the following bound on $|R(z)|$,
\begin{equation}
|R(z)|\leq\sqrt\frac{2}{\pi z}+|C_o|+\frac{2}{\pi}|log z|.
\end{equation}

For small $|z|$ the behavior of the Hankel function gives us
\begin{equation}
|R(z)|\leq C|z|^2[|log z|+1]
\end{equation}

Combining (C.15) and (C.16) we get for real $z>0$,
\begin{equation}
|R(z)|<C~log(1+|z|), ~\forall z>0.
\end{equation}
Next we want to prove that, for $u>0$ and $v>0$,
\begin{equation}
|R(u)-R(v)|<C~log[1+|u-v|].
\end{equation}
We do this in two steps:\\

First we prove that
\begin{equation}
|R(u)-R(v)|<Const~|u-v|,
\end{equation}
then we prove that
\begin{equation}
|R(u)-R(v)|<C_1+C_2~log(1+|u-v|).
\end{equation}
On taking the best of (C.19) and (C.20), and using the fact that
$log(1+x)>\frac{x~log(1+x_0)}{x_0}$ for $0<x<x_0$, (C.18) follows.

To prove (C.19) we proceed as follows:

We have, assuming, without loss of generality, $v>u>0$,
\begin{eqnarray}
|R(v)&-&R(u)|=|H_0^{(1)}(u)-H^{(1)}_0(v)+\frac{i}{2\pi}log\frac{v}{u}|\\ \nonumber
&\leq&\int^v_u dx|H^{(1)}_1(x)+\frac{i}{2\pi x}|dx,
\end{eqnarray}
where we used the property $\frac{d H^{(1)}_0(x)}{dx}=-H^{(1)}_1(x)$.

The expansion of $H^{(1)}_1(x)$ near $x=0$ is
\begin{equation}
H^{(1)}_1(x)=J_1(x)+i[\frac{2}{\pi}J_1(x)log\frac{\gamma x}{2}-\frac{x}{2\pi}\sum\limits_{\ell=o}^{\infty}c_{\ell}x^{2\ell}-\frac{1}{2\pi x}],
\end{equation}
where $\sum c_{\ell}x^{2\ell}$ is an entire function with $c_o=1$.  We thus get
\begin{equation}
|H^{(1)}_1(x)+\frac{i}{2\pi x}|<A x+B x|log x|, ~x<1.
\end{equation}

On the other hand we have also the bound
\begin{equation}
|H^{(1)}_1(x)|<C[\frac{1}{x}+\frac{1}{\sqrt{x}}],
\end{equation}
which follows from Nicholson's formula for $\nu=1$.  Combining (C.24) and (C-.25) we get
\begin{equation}
|H^{(1)}_1(x)+\frac{i}{2\pi x}|\leq Const., \forall x>0.
\end{equation}
Thus, by integration of (C.21)
\begin{equation}
|R(|u|)-R(|v|)|<Const. |u-v|.
\end{equation}

Finally, we need to prove (C.20).  Here, we distinguish three cases with $v>u>0$,
i) ~$v>2, u>1$.

From Eq. (C.21) and from the bound (A.2) we obtain
\begin{eqnarray}
|R(u)-R(v)|&\leq& Const.+\frac{1}{2\pi}~log(\frac{v}{u})\\ \nonumber
&\leq&Const.+\frac{1}{2\pi}~log(1+v-u)
\end{eqnarray}
ii)~$v>2,~u<1$
\begin{equation}
|R(u)-R(v)|\leq C_1+C_2~log(1+v),
\end{equation}
which follows from (C.17).  Hence we have
\begin{equation}
|R(u)-R(v)|\leq C_1+C_2~log(1+v-u)\times~\frac{log 3}{log 2}.
\end{equation}
iii)~ $v<2$.

Here we get from (C.26)
\begin{equation}
|R(|u|)-R(|v|)|\leq C_1+C_2~log(1+v-u)\times\frac{2}{log 2}
\end{equation}

Finally, we stress that (C-.19) is exactly sufficient to establish the uniform continuity of the full
kernel of the Lippmann Schwinger equation given in Eq. (6.1).  The uniform continuity for
$log|\vec{x}-\vec{y}|V(\vec{y})$ has already been established in the main text.  It remains to prove
that $\Delta_R(k;\vec{x},\vec{y})$ operating on any $\chi\varepsilon \it{C}$ leads to a
$(\Delta_R\chi)(\vec{x})$ which is uniformly continuous in $\vec{x}$.  From (C.19) we have,
\begin{equation}
|[R(k|\vec{x}|)-R(k|\vec{x}-\vec{y}|)]-[R(k|\vec{x'}|)-R(k|\vec{x'}-\vec{y}|)]|\leq Const. k|\vec{x}-\vec{x'}|].
\end{equation}
which is what we need.

\vspace{.25in}
\section*{Appendix D}\def\thesection{D}\setcounter{equation}{0}

In this appendix we give a proof of T.
theorem 5.1 stated in section V.

This theorem starts with the case where the zero energy homogeneous integral equation,
\begin{equation}
\phi\vec{(x)}=\frac{1}{2\pi}\int d^2y~log|\vec{x}-\vec{y}|~V(\vec{y})~\phi(\vec{y}),
\end{equation}
has $N$ non-trivial solutions, $\phi_j(\vec{x}), j=1, ..., N,$ which all satisfy the condition
\begin{equation}
\int d^2 xV(\vec{x})\phi_j(\vec{x})=0.
\end{equation}
It then follows that the inhomogeneous integral equation,
\begin{equation}
\phi(\vec{x})= 1+\frac{1}{2\pi}\int d^2y~log|\vec{x}-\vec{y}|V(\vec{y})\phi(\vec{y}),
\end{equation}
has a solution, $\phi_a$.  This solution is of course not unique.

To establish (D-.3) we first show that (D.1) and (D.2) imply that
\begin{equation}
\int d^2x V(\vec{x})[\phi_j(\vec{x})]^2\neq 0.
\end{equation}
This result will be proved at the end of this Appendix.

The number $N$, of linearly independent solutions of $(D.1)$ is finite.  This follows from Fredholm
theory and the compactness of the operator $K$.

We use the notation
\begin{equation}
\phi_o0\vec{x})=\Sigma_{j=1}^N c_j\phi_j(\vec{x}).
\end{equation}

Next we generalize Eq. (D.3) slightly replacing 1 by $f(x)$,
\begin{equation}
\phi(\vec{x})=f(\vec{x})+\frac{1}{2\pi}\int d^2 y~log|\vec{x}-\vec{y}|V(\vec{y})\phi(\vec{y}).
\end{equation}
We multiply by $\phi_0(\vec{x})V(\vec{x})$ and integrate over $\vec{x}$:
\begin{eqnarray}
\int\phi_0(\vec{x})V(\vec{x})\phi(\vec{x})d^2x&-&\int\phi_0(\vec{x})V(\vec{x})f(\vec{x})\\ \nonumber
&=&\frac{1}{2\pi}\int d^2x\int d^2y\phi_0(\vec{x})V(\vec{x})~log|\vec{x}-\vec{y}|V(\vec{y})\phi(\vec{y})\\ \nonumber
&=&\frac{1}{2\pi} \int d^2y\phi(\vec{y})V(\vec{y})\phi_0(\vec{y}).
\end{eqnarray}
Therefore if (D.1) has a solution
\begin{equation}
\int d^2x f(\vec{x})V(\vec{x})\phi_0(\vec{x})=0.
\end{equation}
In other words, (D.2) is a necessary condition.

We proceed to show that (D.2) is also sufficient.  Of course, (D.2) is satisfied for all
$\phi_0(\vec{x})$, and for this purpose, we have to be more careful.

Equation (D-6) is defined on a Banach space, $\bf{B}$.  Thus
\begin{equation}
f\varepsilon \it\bf{B}, ~~and ~~\phi\epsilon\it\bf{B}.
\end{equation}
We define
\begin{equation}
(K\phi)(\vec{x})=\frac{1}{2\pi}\int d^2y~log|\vec{x}-\vec{y}|V(\vec{y})\phi(\vec{y}).
\end{equation}
The Banach space $\bf{B}$ is chosen such that $K$ is a compact operator from $\bf{B}$ to $\bf{B}$.  Thus we
write (D.1) and (D.6) as
\begin{equation}
\phi = K \phi,
\end{equation}
\begin{equation}
\phi = f +K\phi.
\end{equation}

Starting with $\bf{B}$, we want to define a second Banach space as
follows.  The elements of $\bf{B_1}$ are the equivalence classes,
$\{f+\sum^N_{j=1} c_j\phi_j\}$ in $\bf{B}$, where $c_j$ run over
all real numbers.  We denote this equivalence class by $F$, and
$F\epsilon \bf{B_1}$.  We verify that $\bf{B_1}$ is indeed a
Banach space under the norm
\begin{equation}
||F||_1 =\min_{c_j}||f+\Sigma^N_{j=1} c_j\phi_j||.
\end{equation}

Because $\phi_j$ satisfy (D.1), $K$ is also an operator from $\bf{B}_1$ to $\bf{B}_1$.  The point is
\begin{equation}
K(f+\Sigma c_j\phi_j)=K f+\Sigma^N_{j=1}c_j K\phi_j=K f+\Sigma^N_{j=1}c_j\phi_j,
\end{equation}
which satisfies the definition of being an equivalence class.  In the second Banach space $\bf{B}_1$, (D.3) takes the form
\begin{equation}
\Phi = ~F~+~K~\Phi.
\end{equation}
We verify that $K$ is a compact operator from $\bf{B}_1$ to $\bf{B}_1$.  Thus (D.15) is a Fredholm
equation in $\bf{B}_1$.

We apply the Fredholm alternative to (D.15).

The Banach space $\bf{B}_1$ is constructed such that the homogeneous equation,
\begin{equation}
\Phi\ = ~K~\Phi
\end{equation}
has no non-trivial solutions in $\bf{B}_1$.  To show this we assume the contrary, i.e. that there is a non-trivial $\Phi$.
Translated back to the original Banach space, $\bf{B}$, (D.16) is
\begin{equation}
\phi =\phi_0 +~K\phi
\end{equation}
where $\phi_0$ satisfies (D.1) and (D.2).  By (D.8), the existence of a solution implies that
\begin{equation}
\int\phi_0(\vec{x})V(\vec{x})\phi_0(\vec{x})d^2x=0,
\end{equation}
i.e.
\begin{equation}
\int d^2xV(\vec{x})[\phi_0(\vec{x})]^2=0.\\ \nonumber
\end{equation}
But this contradicts (D.54).  Therefore (D.16) cannot have a non-trivial solution.

From the Fredholm alternative, we know that (D.15) always has a solution in the Banach space $\bf{B_1}$.
Translated back to the original space $\bf{B}$, this means that
\begin{equation}
\phi=f+\Sigma^N_{j=1}c_j\phi_j+K\phi,
\end{equation}
has a solution in $\bf{B}$ for a suitably chosen set of coefficients $c_j$.

It therefore remains to show that
\begin{equation}
c_j = 0,~~ ~~j=1, ... ,N.
\end{equation}
To prove this we again assume the opposite, i.e. not all $c_j$ are
zero, or $\phi_0=\Sigma^N_{j=1} c_j\phi_j$ is non-trivial.  Then
we have both
\begin{equation}
\int[f(\vec{x})+\phi_0(\vec{x})]V(\vec{x})\phi_0(\vec{x})d^2x=0,
\end{equation}
and
\begin{equation}
\int f(\vec{x})V(\vec{x})\phi_0(\vec{x})d^2x=0.
\end{equation}
The latter is just (D.8).  Subtracting these last two equations we obtain
\begin{equation}
\int V(\vec{x})[\phi_0(\vec{x})]^2d^2x=0.
\end{equation}
But this again contradicts (D-.4).

The conclusion is therefore reached that the necessary and sufficient condition for (D.6) to have a solution
is (D.8), or
\begin{equation}
\int f(\vec{x})V(\vec{x})\phi_j(\vec{x})d^2x=0
\end{equation}
for $j=1, ... ,N$.

It only remains to specialize (D.6) to the case $f=1$.  Thus the necessary and sufficient condition for (D.3)
to have a solution is $\int d^2x V(\vec{x})\phi_j(\vec{x})=0, ~~j=1, ... ,N$.

We are only left with the task of proving (D.4),i.e. that $\int V(\vec{x})[\phi_0(\vec{x})]^2d^2x\neq 0.$

A zero energy bound state is characterized by the a solution, $\phi_0(\vec{x})$, of the equation
\begin{equation}
\phi_0(\vec{x})=\frac{1}{2\pi}\int d^2y ~log~ k_0|\vec{x}-\vec{y}|V(\vec{y})\phi_o(\vec{y}).
\end{equation}
with the condition
\begin{equation}
\int d^2x V(\vec{x})\phi_0(\vec{x})=0
\end{equation}

From (D.26) we see that (D.27) is independent of the scale $k_0$.  We set $k_0=1$.

Using the Schrodinger equation,
\begin{equation}
-\nabla^2\phi_0+V(\vec{x})\phi_0 =0.
\end{equation}
we get, after multiplication by $\phi_0$ and integration
\begin{equation}
-\int\limits_{\vec{|x|}\leq
R}d^2x\phi_0\nabla^2\phi_0+\int\limits_{|\vec{x}|*\leq R}d^2x
V(\vec{|x|})\phi_0(\vec{x})^2=0.
\end{equation}
Integrating by parts we get,
\begin{equation}
F(R)=\int\limits_{|\vec{x}|<R}|\nabla\phi_0|^2d^2x+\int\limits_{|\vec{x}|<R} V(\vec{x})\phi^2_0(\vec{x})d^2x=
\int\limits_{|\vec{x}|=R}ds.(\vec{\nabla}\phi_0)\phi_0.
\end{equation}
Using polar coordinates we have
\begin{equation}
\frac{F(R)}{R}=\int d\theta\phi_o(R_1)\frac{\partial\phi_o(r,\theta)}{\partial r}|_{r=R}.
\end{equation}
Taking $R_1\leq R\leq R_2$ we obtain
\begin{equation}
\int^{R_2}_{R_1}\frac{F(R)}{R}~dR=\frac{1}{2}\int d\theta[\phi^2_o(R_2,\theta)-\phi^2_o(R_1,\theta)].
\end{equation}

In Appendix E we prove that if (D.26) and (D.27) hold, $\int
d\theta\phi^2_0(R,\theta)\rightarrow 0$ as $R\rightarrow\infty$.

If this result is true, we then have
\begin{equation}
\int^{R_2}_{R_1}\frac{F(R)}{R}~dR\leq Const., ~~\forall~R_2>R_1>>R_o.
\end{equation}

Hence using the mean value theorem, $\exists$ an $\overline{R}$, $R_1<\overline{R}<R_2$, such that
\begin{equation}
F(\overline{R})\leq\frac{C}{log(R_2/R_1)}.
\end{equation}
Therefore there is a sequence, $\overline{R}_1,\overline{R}_2, ... ,\overline{R}_j$, such that
$F(\overline{R}_j)\rightarrow 0$ as $j\rightarrow\infty$.

But $\int d^2x|V|[\phi_0(\vec{x})]^2$ is convergent since $|\phi_o|<Const. log(2+|\vec{x}|)$ for large $|\vec{x}|$.
Potentials with $\int d^2x |V|[log(2+|\vec{x}|)]^2<\infty$ belong to our class.  Hence we conclude that
$\int\limits_{|\vec{x}|<R}V\phi^2_0d^2x$ has a limit as $R\rightarrow\infty$.

On the other hand the first term on the $r.h.s.$ of (D.29), i.e. $\int\limits_{|\vec{x}|<R}|\vec{\nabla}
\phi_0|^2d^2x$, is a monotonically increasing function of $R$, so it either has a limit as $R\rightarrow\infty$
or it tends to $+\infty$.  The latter case is in contradiction with the fact that
$F(\overline{R}_j)\rightarrow 0$ as $j\rightarrow\infty$.  Therefore $\int\limits_{|\vec{x}|\leq R}d^2x
\vec{|\nabla}\phi|^2$ has a limit as $R\rightarrow\infty$, and $F(R)$ has a limit which is
identically zero.  But, $\int d^2x|\vec{\nabla}\phi_o|^2>0$ strictly, and therefore
\begin{equation}
\int V[\phi_0]^2d^2x<0.
\end{equation}

\vspace{.25in}
\section*{Appendix E}\def\thesection{E}\setcounter{equation}{0}

In this Appendix we prove the following:  Given
$\phi(\vec{x})\epsilon \cal{B}$ which satisfies the integral
equation
\begin{equation}
\phi(\vec{x})=c_0+\frac{1}{2\pi}\int
d^2y(log|\vec{x}-\vec{y}|)V(\vec{y})\phi(\vec{y}),
\end{equation}
with $c_0$ finite or zero, then if
\begin{equation}
\int d^2xV(\vec{x})\phi(\vec{x})=0,
\end{equation}
$|\phi(\vec{x})|$ is uniformly bounded for all $\vec{x}\epsilon
R_2$.

We stress first that, as in Appendix D, the $\phi(\vec{x})$
considered here are all in the Banach space $\cal{B}$, and hence
$\parallel\phi\parallel_{\ell}\leq Const.$.  This means that we
have ab initio the bound
\begin{equation}
|\phi(\vec{x})|\leq C
log(2+|\vec{x}|).
\end{equation}
Thus to complete the task of this Appendix we only have to study
the large $|\vec{x}|$ behavior of $|\phi|$.

Using again the notation $log A=log^+A-log^-A$, we set
\begin{equation}
\frac{1}{2\pi}\int d^2y(log|\vec{x}-\vec{y}|)V(\vec{y})\equiv
I^+(\vec{x})-I^-(\vec{x}).
\end{equation}

First we prove that
\begin{equation}
I^+(\vec{x})\longrightarrow 0,
~~as~~|\vec{x}|\longrightarrow\infty.
\end{equation}

Given (E.2) we can write $I^+$ as
\begin{equation}
I^+(\vec{x})=\frac{1}{2\pi}\int d^2
y[log^+|\vec{x}-\vec{y}|-log(2+|\vec{x}|)]V(\vec{y})\phi(\vec{y}).
\end{equation}

We define $y_0(x)$ as
\begin{equation}
y_0\equiv\frac{x}{log(2+x)}.
\end{equation}
Next we split the integration in (E.6)
\begin{eqnarray}
|I^+|&\leq&\frac{1}{2\pi}\int\limits_{y<y_0(x)} d^2y
|log|\vec{x}-\vec{y}|-log(2+x)| |V(\vec{y})\phi(\vec{y})|\\
\nonumber &+&\frac{1}{2\pi}\int\limits_{y>y_0(x)} d^2y[2log(2+x)
+log(2+y)] |V(\vec{y})\phi(\vec{y}).
\end{eqnarray}
In the first integral, for $x$ large enough,
\begin{equation}
|log|\vec{x}-\vec{y}|-log(2+x)|<\frac{2}{log(2+x)},
\end{equation}
But as~$x\rightarrow\infty$,
\begin{equation}
\int\limits_{y<y_o(x)}
d^2y|log|\vec{x}-\vec{y}|-log(2+x)||V(\vec{y})||\phi(\vec{y})|
\leq\frac{C}{log(2+x)}\int\limits_{y<y_0(x)}d^2y|V(y)|log(2+y).
\end{equation}
Hence the first integral vanishes as $x\rightarrow \infty$.  The
second integral, $I^+_2$, satisfies
\begin{equation}
|I^+_2(\vec{x})|\leq
C\int\limits_{y>y_0(x)}d^2y|V(y)|[log(2+y)]^2=o(1),
\end{equation}
as $x\rightarrow \infty$.  Thus Eq. (E.5) is proved.

At this stage we have
\begin{equation}
|\phi(\vec{x})|\leq C+\frac{1}{2\pi}\int
d^2y(log^-|\vec{x}-\vec{y}|)|V(\vec{y})| |\phi(\vec{y})|
\end{equation}

Next we define $M(r)$ and  $B(r)$ as follows:
\begin{equation}
M(r)=\sup\limits_{|\vec{x}|\leq r}|\phi(\vec{x})|,
\end{equation}
\begin{equation}
B(r)=r \sup\limits_{\rho\leq r}\frac{M(\rho)}{\rho}\geq M(r).
\end{equation}

$B(r)$ exists because $M(r)\leq C log(2+r)$.  If $|\phi(\vec{x})|$
grows to infinity as $|\vec{x}|\rightarrow\infty$, there must
exist a sequence
$\vec{x}_1,\vec{x}_2,...,\vec{x}_n,|\vec{x}_n|\rightarrow\infty$
as $n\rightarrow\infty$, where $M(|\vec{x}_n|)=|\phi(\vec{x}_n)|$,
and there must be a subsequence
$\vec{x}_1',\vec{x}'_2,...,\vec{x}'_N$, where
\begin{equation}
B(|\vec{x}'_N|)=M(|\vec{x}'_N|)=|\phi(\vec{x}'_N)|.
\end{equation}
We take such an $\vec{x}'_N$, and obtain
\begin{equation}
B(|\vec{x}'_N|)\leq C+\frac{1}{2\pi}\int
d^2y(log^-|\vec{x}'_N-\vec{y}|)|V(\vec{y})|.B(|\vec{x}'_N|)(\frac{|\vec{x}'_N|+1}{|\vec{x}'_N|})
\end{equation}
where we recall that in the integral
\begin{equation}
|\vec{x}'_N|-1\leq|\vec{y}|\leq |\vec{x}'_N|+1.
\end{equation}

Now we obtain
\begin{eqnarray}
\int
d^2ylog^-|\vec{x}'_N-\vec{y}||V(\vec{y})|&\leq&
\int\limits_{|\vec{z}|<\Delta}(log^-|\vec{z}|)|V(\vec{z})|_Rd^2z\\
\nonumber &+&|log
\Delta|\int\limits_{|\vec{y}|>|\vec{x}'_N|-1}|V(\vec{y})|d^2y.
\end{eqnarray}

The first integral can be made arbitrarily small by taking
$\Delta$ small enough.  Once $\Delta$ is fixed we can make the
second integral as small as we please by taking $|\vec{x}'_N|$
large enough.  Hence we finally have
\begin{equation}
B(|\vec{x}'_N|)\leq C+\epsilon B(|\vec{x}'_N|),
\end{equation}
with $\epsilon <1/2$ for $|\vec{x}'_N|$ large enough.

Therefore $B$ is bounded, and it follows that $|\phi|$ is bounded.

From this it is easy to see that
\begin{equation}
\int
d^2y(log^-|\vec{x}-\vec{y}|)|V(\vec{y})||\phi(\vec{y})|\rightarrow
0,
\end{equation}
as $|\vec{x}|\rightarrow\infty$, because from (E.18),
\begin{equation}
\int d^2y(log^-|\vec{x}-\vec{y}|)|V(\vec{y})|\rightarrow
0~~as~~|\vec{x}|\rightarrow\infty.
\end{equation}

As a consequence, if in Eq. (E.1), $C_0=0$,then $|\phi|\rightarrow
0$ as $|\vec{x}|\rightarrow\infty$.

In the case where $\int d^2xV(\vec{x})\phi(\vec{x})\neq 0$, we
have
\begin{equation}
|\int dy^2 log^-|\vec{x}-\vec{y}| V(\vec{y})\phi(\vec{y})|\leq C
log(2+x)\int d^2y log^-|\vec{x}-\vec{y}||V(\vec{y})|.
\end{equation}

Combining this with (E.2) and using Eq. (E.21) we get, for
$|\vec{x}|\rightarrow\infty$
\begin{equation}
|\phi|\cong \frac{1}{2\pi}log(2+x)|\int
d^2yV(\vec{y})\phi(\vec{y})+o(1)|.
\end{equation}.
\section*{Acknowledgements}
\vspace{.25in}
The authors are indebted to Henri Epstein for several helpful discussions.  In addition he provided
a much more elegant and general proof for the compactness theorem given in Appendix B.  We also wish
to thank K. Chadan, B. Morariu and H.-c. Ren for helpful remarks.  Two of us, N.N.K. and T.T. Wu, are
grateful to the CERN Theory Division for its kind hospitality.  One of us, A. Martin, also thanks the
theory group at Rockefeller University for its hospitality and support.  This work was supported in part by the
U.S. Department of Energy under Grant No. DE-FG02-91ER40651m Task B, and under Grant No. DE-FG02-84ER40158.

\section*{References}
\vspace{.25in}
1.  K. Chadan, N.N. Khuri, A. Martin and T.T.Wu, Phys. Rev. \underline{D58}, 025014 (l998).
2.  D. Boll\'{e} and F. Geztesy, Phys. Rev. Lett. \underline{52}, 1469 (l984).\\
3.  J. Bros and D. Iaglonitzer, Phys. Rev. D,\underline{27}, 811) (l983).\\
4.  F. Dalforo, S. Giorgini, L.P. Pitaevskii and S. Stringari, Rev. of Mod. Phys., \underline {71},
    463 (1999).\\
5.  E.H. Lieb and J. Yngvasson, J. Stat. Phys., \underline{103}, 509 (2001).\\
6.  H.-c. Ren, "The Virial Expansion of a Dilute Bose Gas in Two Dimensions," cond-mat/0307342,
    J. of Stat. Phys., in press.\\
7.  J. Dieudonne, "Foundations of Modern Analysis," (New York, academic Press 1960) page 137.
8.  K. Chadan, N.N. Khuri, A. Martin and T.T. Wu, J. of Math. Physics,.\\
9.  A Martin and T,T, Wu, preprint CERN-TH/2003-235 (2003).\\
10. W. Magnus and F. Oberhettinger, "Formulas and Theorems for The Functions of Mathematical Physics,"
    Chelsea Publishing Co., New York, 1954), page 31.\\
11. ibid, page 33.\\
12. M. Abramovitz and I. Stegun, "Handbook of Mathematical Functions,"  (Dover Publications, New York,
    1972), page 305, eqn. (9.2.25).

\end{document}